\documentclass[12pt,oneside]{dissertation}
\usepackage{float}
\restylefloat{figure}
\usepackage{comment}
\usepackage{subfig}

\author{Justin R. Mason}
\title{Quantum Corrections to Diffusion in Stars}
\degree{Master of Arts}
\degreedate{August 2011}
\department{Astronomy}

\defensedate{August 23, 2011}

\begin{document}
	\setcounter{chapter}{0}

	\pagenumbering{roman}

	\maketitle

	\begin{acceptance}
		\sigline{Dr. Charles Horowitz}
		\sigline{Dr. Thomas Steiman-Cameron}
		\sigline{Dr. Constantine Deliyannis}
	\end{acceptance}

	\copyrightpage{2011}



	\begin{abstract}
\begin{singlespace}
Quantum corrections can be important for diffusion and the melting temperature of dense plasmas in compact astrophysical objects, particulary white dwarfs and neutron stars. Typically ions in these systems are modeled classically, but Daligault \emph{et al}. use a semiclassical inter-ion potential. We run molecular dynamic simulations using this semiclassical approach in order to calculate the diffusion coefficient and melting temperatures in a one component plasma. We find that in liquid simulations quantum corrections do not have a significant effect on diffusion, increasing it by only a factor of two from our classical simulation. However, in solid simulations, diffusion slowly increases for small quantum corrections, but once quantum effects become large enough, the system can liquify. We also find that quantum corrections can decrease the melting temperature of a one component plasma, potentially affecting the way in which white dwarfs cool. These results suggest that a helium white dwarf may remain liquid at typical white dwarf densities, while the structure of a neutron star's crust could be altered due to quantum effects.
\end{singlespace}
		\vspace{0.50in}
		\sigline{Dr. Charles Horowitz}
		\vspace{-0.50in}
		\sigline{Dr. Thomas Steiman-Cameron}
		\vspace{-0.50in}
		\sigline{Dr. Constantine Deliyannis}
		\vspace{-0.50in}
	\end{abstract}
	
	\tableofcontents
	\newpage

	\listoftables
	\newpage

	\listoffigures
	\newpage

	\pagenumbering{arabic}

	\chapter{Introduction}
\label{chap:introduction}

\indent \indent By understanding quantum effects in diffusion in white dwarfs (WD) and neutron stars (NS), one can determine more accurately the structure, age, and evolution of the Milky Way Galaxy. In a WD, sedimentation of ions with a large mass-to-charge ratio releases gravitational energy and slows their cooling rates \cite{Winget 2009}. The crust of a NS undergoes chemical separation during crystallization at extreme densities \cite{Horowitz 2007}. This separation changes physical properties of the crust. Diffusion constants for a system can be determined from Molecular Dynamics (MD) simulations. As computational power has increased a lot of effort has been put into MD simulations to determine diffusion constants for these Coulomb plasma systems. These simulations are extremely powerful tools that can be used to understand the underlying physics for systems such as WDs or NSs by matching observational evidence.

\indent WD stars are the remnants of low- and intermediate-mass stars that can no longer sustain nuclear fusion and are believed to mark the end of stellar evolution for more than 97\% of stars \cite{Fontaine 2001}. WDs spend their lives slowly cooling by radiating thermal energy from their cores \cite{Kippenhahn 1994}. WDs are extremely compact and dense stellar objects. Their mass is on the order of that of the Sun, but they are approximately the size of the Earth. Reaching $n\sim$10$^{28}$-10$^{30}$ cm$^{-3}$ densities, they are comprised of a pressure ionized Coulomb plasma, which is stopped from further collapse by electron degeneracy pressure. In such a degenerate gas the electrons are distributed fairly uniformly surrounding the nulcei. However, due to strong Coulomb interactions at high density the ions undergo a phase transition to form a plasma crystal. Any energy released due to gravitational collapse is used in raising the Fermi energy of the electrons (forcing degenerate electrons into higher energy levels) rather than increasing the star's luminosity \cite{Kippenhahn 1994}.

\indent The WD luminosity function is the number of WD stars observed as a function of magnitude. Structure within the WD luminosity function (LF) provides information on the age and timescale of formation for components of the Milky Way Galaxy. Interpretations of the WD LF structure is dependent upon our underlying knowledge of cooling physics \cite{Fontaine 2001}. In the field of WD cosmochronology, these slowly changing stars are used to track the age and star formation history of the Milky Way. A good review of WD cosmochoronology is given by Fontaine \emph{et al.} \cite{Fontaine 2001}. Precise determinations of the WD LF in globular cluster NGC 6397 has shown a peak around a magnitude of 26.5. A build up of WDs at this magnitude is attributed to the release of latent heat of crystallization that delays the rate at which WDs cool \cite{Winget 2009}. While observations of WD cooling in NGC 6791, a metal rich open cluster, shows that the WD luminosity function is well fit by evolutionary models only when including both sedimentation of neutron rich ions such as $^{22}$Ne, which releases gravitational energy, and the release of latent heat of crystallization \cite{Garcia 2010, Winget 2009}. Modeling diffusion precisely in WDs allows one to estimate the rate of sedimentation and thus determine more accurately the age of a stellar cluster.

\indent While most stars will end as WDs, high-mass stars end as NSs. As a high mass star reaches the end of its lifespan its core can reach densities too high to be supported by electron degeneracy pressure. The electron Fermi energy increases to a point where inverse beta decay occurs and electrons are forced into the nuclei,
\begin{equation}
p + e^- \rightarrow n + \nu.
\end{equation} 
\noindent It is well known that free neutrons are unstable and will beta decay with a half-life of just under 15 minutes. However, this reaction is not allowed to happen within a NS because the electron Fermi energy is too high. The neutrons in a NS are stable because of their environment \cite{Kippenhahn 1994}.

\indent The typical size of a NS is approximately 12 km with a mass between 1.4-2.0 M$_{\odot}$. The maximum mass, size, and interior structure of a NS is very equation of state (EOS) dependent. The mechanism keeping a NS from further collapse is neutron degeneracy pressure and strong interactions.  The outer crust of a neutron star is comparable in density to the interior of a WD and can have a complicated structure. If a NS continually accretes material from a companion star the infalling material can undergo nuclear reactions by rapidly capturing protons (the rp process). The material also increases the density of the liquid ocean layer until it crystallizes. The result is a crust of complex structure that is believed to have a top liquid ocean layer of low-Z nuclei with a bottom solid crust of high-Z nuclei.

\indent Phase separation of high- and low-Z ions in the liquid and solid crusts has been described and explicitly modeled with MD simulations by Horowitz \emph{et al.} 2007 \cite{Horowitz 2007}. Chemical separation between the liquid and solid crust is believed to change properties of the crust and impact many observables: thickness, shear modulus, and breaking strain of the crust. Possible observable effects include changes in the shape of a NS, radiation of gravitational waves, and properties of quasiperiodic oscillations observed in magnetar giant flares \cite{Horowitz 2007}.

\indent Diffusion in Coulomb plasmas in the liquid phase has been well studied since the 1970s. Hansen \emph{et al.} performed MD simulations for the one component plasma (OCP) which consists of ions interacting with pure Coulomb interactions and an inert neutralizing background charge density \cite{Hansen 1975}. More recent work has been done to understand diffusion in Coulomb crystals \cite{Hughto 2011}. Particles exchanging lattice sites or the diffusion of imperfections in the crystal can be important mechanisms for the release of gravitational energy. In simulations, the diffusion in Coulomb crystals is dependent on the form of the inter-particle potential. The Lennard Jones potential with its hard-core contribution ($r^{-12}$) tends to form a glass \cite{Faller 2003} with low diffusion rates while the classical Coulomb (1/r) potential allows ions to diffuse much more easily.

\indent Another commonly used interaction is that of the Yukawa potential. Diffusion for a Yukawa fluid has been simulated by Robbins \emph{et al.} \cite{Robbins 1987} and Ohta \emph{et al.} \cite{Ohta 2000}. In a Yukawa fluid ions interact via a screened Coulomb potential \emph{$v_{ij}$}$(r)$,
\begin{equation}
v_{ij}(r) = \frac{Z_iZ_je^2}{r}e^{-r/\lambda}, \label{eq1}
\end{equation}
for two ions with charges $Z_i$ and $Z_j$ that are separated by a distance r. The OCP is equivalent to a Yukawa fluid, where all of the ions have the same charge ($Z_i=Z_j$), and the Thomas Fermi screening length $\lambda$ is very large. 

\indent In WDs and NSs the motions of ions are typically considered to be classical because of their large masses. However, at extreme densities, quantum corrections have been considered \cite{Daligault 2005,Chabrier 1992,Nag 2000}. In 2005, Daligault \emph{et al.} studied quantum corrections in diffusion by modifying the interaction $v_{ij}(r)$ in a semiclassical approximation which includes the effects of zero point motion. It is the intent of this research to apply this semiclassical approximation to MD simulations of diffusion in crystals and to compare with work in which a purely classical approach has been used.

\indent In the interiors of WDs and the crusts of NSs the inter-ionic spacing, which is close to the size of the ion-sphere radius $a=(3/4\pi n)^{1/3}$ where $n$ is the ion density, becomes comparable to the ionic thermal deBroglie wavelength $\Lambda_{th}=\sqrt{2\pi\hbar^2/MT}$, where M is the ionic mass \cite{Chabrier 1992}. As these values become comparable quantum effects become more important.

\indent We choose to define the parameter $\Lambda=\Lambda_{th}$/$\sqrt{2\pi^2}$ following ref. \cite{Daligault 2005}. The dimensionless quantity $\Lambda/a$ can be used as a measure of the importance of quantum effects in a system. For systems in which $\Lambda/a$ $\ll$ 1 interactions can be considered to be classical. For those where $\Lambda/a$ $\sim$ 1 quantum effects are extremely important. For WDs and NSs they lie between the two previous cases with $0\lesssim \Lambda/a\lesssim$ 1 and a semiclassical approximation to particle interactions can be used. The semiclassical addition to the ionic potential is given as a variation of the Yukawa potential and takes the form
\begin{equation}
v_{ij}(r) = \frac{Z_iZ_je^2}{r}e^{-r/\lambda}(1-e^{-r/\Lambda}). \label{eq2}
\end{equation}
This form for the interaction accounts for the extension of the particle by effectively smearing out the ion on the scale of $\Lambda$. When $\Lambda$ is small this form reduces to that of Eq. \eqref{eq1}. Such a form for the interactions makes the ions softer-core and should allow diffusion to more readily occur for larger values of $\Lambda/a$. Larger values of $\Lambda/a$ can be achieved for lighter elements (eg. $^{4}$He), colder temperatures, or higher densities. These effects are explored in more detail in chapter 4.

\indent In this paper we present our MD formalism in chapter \ref{chap:formalism}. Diffusion constants and melting temperatures for classical and semiclassical simulations are presented in chapter \ref{chap:results}. In chapter \ref{chap:discussion} we discuss implications of this research on the understanding of structure within WDs and NSs. We conclude in chapter \ref{chap:conclusion}.
	
 	\chapter{Formalism}
\label{chap:formalism}
\graphicspath{{./Figures/}}

\indent\indent An MD simulation is a technique in which a system of interacting ions is evolved with time by integrating the ions' equations of motion. Given an initial set of positions and velocities for all ions within the system, the time evolution is completely determined, in principle. Time evolution is based on Newton's law, $F=ma$, and the forces are obtained as the gradient of a chosen potential, which is a function of all the particle coordinates. Positions and velocities of ions are evolved through the velocity Verlet algorithm \cite{Verlet 1967}. A detailed description of many aspects of MD simulations is given by Ercolessi \cite{Ercolessi}.

\section{\label{sec:MDsim}Molecular Dynamics Simulations}
\indent \indent For the purposes of this research, we consider only a one component plasma (OCP). A star with near solar metallicity which has converted most of its original carbon, nitrogen, and oxygen into $^{22}$Ne, might have of order 2\% $^{22}$Ne. The ratio of carbon to oxygen in the core depends on the rates for the $^{4}$He(2$\alpha$,$\gamma$)$^{12}$C and $^{12}$C($\alpha$,$\gamma$)$^{16}$O reactions and is expected to be near one to one \cite{Horowitz 2010}. These conditions can lead to a WD with a core composition close to 49\% $^{12}$C, 49\% $^{16}$O, and 2\% $^{22}$Ne. Because of this high concentration of $^{16}$O, we consider a OCP composed entirely of $^{16}$O. Later simulations may consider using a multi-component plasma to compare with previous work done by Hughto $\emph{et al.}$ \cite{Hughto 2010}.

\indent Ions interact according to Eq. \eqref{eq2}. The Thomas Fermi screening length $\lambda$, for relativistic electrons, is $\lambda^{-1}=2\alpha^{1/2}\emph{k}_{F}/\pi^{1/2}$ where $\emph{k}_{F}$ is the electron Fermi momentum  and is given as $\emph{k}_{F}=(3\pi^{2}\emph{n}_{e})^{1/3}$ and $\alpha$ is the fine structure constant. The electron density, $n_{e}$, is equal to the ion charge density, $n_{e}=Zn$, where $n$ is the ion density and $Z$ is the ion charge. For simplicity, we use the extreme relativisitic limit by neglecting the electron mass. However, this could become more important at lower densities, as this decreases $\lambda$. 

\indent Simulations can be described by a Coulomb coupling parameter 
\begin{equation}
\Gamma =\frac{Z^2e^2}{aT}. \label{eq3}
\end{equation}
Here T is the temperature of the system. $\Gamma$ is the ratio of the Coulomb potential energy to the thermal kinetic energy, and it is known that a OCP crystallizes at $\Gamma\approx 175$. It is also known that the crystallization value of $\Gamma$ may also depend slightly on the value of $\lambda$ \cite{Potekhin 2000}.

\indent One of the most fundamental timescales in plasma physics is that of the plasma frequency $\omega_p$. Long wavelength fluctuations in the charge density can undergo oscillations at the plasma frequency. Therefore, in our simulations, time can be measured in units of $\omega_p^{-1}$.  The plasma frequency depends on the ion charge $Z$ and mass $M$,
\begin{equation}
\omega_p=\left[\frac{4\pi Z^2e^2n}{M}\right]^{1/2}. \label{eq4}
\end{equation}

\indent All simulations are evolved with the velocity Verlet algorithm \cite {Verlet 1967} using time steps of $\Delta t\approx 1/9\omega_p$. Periodic boundary conditions are used in each orthogonal direction. We do not use a cutoff distance for interaction and, therefore, calculate the interactions between all ions. The force on an ion is evaluated by summing over all other ions. In an effort to decrease finite size effects, all simulations use $N=8192$ with a box size $L$ that is much larger than the electron screening length, $L/2=8.92\lambda$. Temperatures are held nearly constant by periodically rescaling the velocities every 2.36/$\omega_p$ (twenty time steps). Any bulk flow of ions is treated by calculating the center of mass velocity in all directions and subtracting from the velocity of each ion every 118/$\omega_p$ (one thousand time steps).

\section{\label{sec:liquiddiff}Liquid Phase Simulations}
\indent \indent In liquids the diffusion constant $D$ can be calculated from the velocity autocorrelation function,
\begin{equation}
Z(t)=\frac{\langle\textbf{v}_j(t_0+t)\cdot\textbf{v}_j(t_0)\rangle}{\langle\textbf{v}_j(t_0)\cdot\textbf{v}_j(t_0)\rangle}. \label{eq5}
\end{equation}
This averages over all ions $j$ and over initial times $t_0$ to minimize statistical errors. The velocity of the $\emph{j}$th ion at time $t$ is $\textbf{v}_j(t)$. The diffusion constant is calculated from the time integral of $Z(t)$,
\begin{equation}
D=\frac{T}{M}\int_0^{t_{max}} dtZ(t). \label{eq6}
\end{equation}
\noindent With time $Z(t)$ fluctuates about zero, and thus contributions to $D$ are significantly reduced by a time $t_{max}=240/\omega_p$.

\indent To ensure that simulations with the semiclassical approximation begin with similar initial conditions, a system with a classical interaction is allowed to equilibrate for a time of $1200/\omega_p$ (10,000 time steps). The ions begin with random initial positions at $\Gamma=150$. Ions are given random initial velocities between $\pm 3kT/M$. As the simulation evolves the ions lose the random velocity distribution and assume one of a Gaussian distribution. The final positions and velocities for this system are written out to a file and used as the initial conditions for subsequent liquid phase simulations. All liquid phase simulations which include quantum corrections use the classical positions and velocities for their initial conditions. A value for $\Lambda$ is then introduced and the systems are allowed to equilibrate for a time of $1200/\omega_p$. Simulations are then to evolved for an additional $1200/\omega_p$ during which the positions and velocities of ions are written to a trajectory file after every time step. Results for $Z(t)$ and $D(t)$ are given in section \ref{liquidres}.

\section{\label{sec:soliddiff}Solid Phase Simulations}
\indent \indent Diffusion in crystals is much smaller than in liquids making it difficult to calculate diffusion coefficients using Eq. \eqref{eq5}. As $Z(t)$ fluctuates about zero the integral in Eq. \eqref{eq6} involves sensitive cancellations of $D(t)$ where $Z(t)$ is positive and negative. A more effective method of calculating $D(t)$ in solids is 
\begin{equation}
D(t)=\frac{\langle|\textbf{r}_j(t+t_0)-\textbf{r}(t_0)|^2\rangle}{6t}, \label{eq7}
\end{equation} 
\noindent where $\textbf{r}_j(t)$ is the position of the $j$th ion at time $t$ and the average is over all ions $j$ and initial times $t_0$. The diffusion constant $D$ is the large time limit of Eq. \eqref{eq7},
\begin{equation}
D=\lim_{t \to \infty} D(t). \label{eq8}
\end{equation}

\indent However, simulations are finite in time, and thus thermal oscillations about lattice sites can contribute to $D(t)$ for small time limits of Eq. \eqref{eq8}. An ion oscillating about a lattice site does not contribute to the net diffusion of the system, but it will have $|\textbf{r}_j(t)-\textbf{r}_j(0)|$ nonzero and contribute to Eq. \eqref{eq8}. Therefore, thermal oscillations cause $D(t)$ to differ from $D$ for small $t$. It is convenient to define the quantity
\begin{equation}
D'(t)=\frac{\langle\Theta[|\textbf{r}_j(t')-\textbf{r}_j(t_0)|-R_c]|\textbf{r}_j(t')-\textbf{r}_j(t_0)|^2\rangle}{6t}, \label {eq9}
\end{equation}
where $t'=t+t_0$. Eq. \eqref{eq9} introduces a cutoff radius $R_c$ for which ions are required to travel before contributing to the net diffusion. The cutoff radius is of order of the lattice spacing and reduces contributions of thermal oscillations to $D'(t)$. It has been observed by Hughto \emph{et al.} \cite{Hughto 2011} that $D'(t)$ is approximately independent of $t$, even for moderate $t$, so that
\begin{equation}
D\approx D'(t). \label{eq10}
\end{equation}

\indent It is also important to note that at arbitrarily large $t$, ions which diffuse distances nearly equivalent to the width of the box  introduce error in Eq. \eqref{eq7} and \eqref{eq8} due to periodic boundary conditions. However, diffusion is relatively slow in solids so this is often not a problem until very large $t$. 

\indent Initial conditions are very important for determining $D$, as systems can possibly contain defects which may take a long time to equilibrate. Hughto \emph{et al.} describe a simulation in \cite{Hughto 2011} that suggests both WD and NS plasmas freeze into nearly perfect body-centerd cubic (bcc) crystals. Because of their result, we create initial conditions from a classically interacting system ($\Lambda /a=0$) of a pure bcc crystal at a temperature equivalent to $\Gamma =300$ and slowly warm it to $\Gamma =175$. Finally, the quantum correction is introduced to the interactions, and the system is allowed to equilibrate for $1200/\omega_p$ and then evolved for $4800/\omega_p$ (40,000 time steps) while the ions' positions and velocities are recorded every $23.6/\omega_p$ (200 time steps).

\indent Diffusion in a crystal at a temperature of $\Gamma =200$ are also considered. The initial conditions begin from the same initial conditions as $\Gamma=175$ simulations. However, when the $\Gamma=200$ simulations are allowed to equilibrate the system is cooled off by adjusting the ions' velocities along with introducing the value for $\Lambda$. Systems are equilibrated for $1200/\omega_p$ and then evolved for $4800/\omega_p$ while the ions' positions and velocities are recorded every $23.6/\omega_p$.

\indent In both temperature regimes $(\Gamma=175, 200)$ Eq. \eqref{eq9} is used to calculate the diffusion constant with $t'\approx 4250 \omega_p$. This allows the diffusion constant to be averaged over twenty configurations. The initial positions $r_j(t_0)$ of each configuration are separated by $23.6/\omega_p$. Final positions $r_j(t')$ are separated from initial conditions by $t\approx 4250/\omega_p$. This is done to decrease statistical uncertainties in $D$. Histograms of number of ions diffused versus displacement are constructed and presented with diffusion results in chapter \ref{sec:solidres}.

\section{\label{sec:tempsim}Melting Temperature}
\indent \indent Another aspect in which this research can provide insight is that of the melting temperature of a system that includes quantum corrections, $v_{ij}$ equal to that of Eq. \eqref{eq2}. It is known that the melting temperature of a classically interacting system is $\Gamma\approx175$ \cite{Potekhin 2000}, but little work has been done for systems which include quantum corrections.

\indent The equilibrium temperature is found by evolving a system of half liquid and half solid then visually checking if the system begins to favor one configuration. If the system is out of equilibrium it will begin to liquify or crystallize accordingly, and the value of $\Gamma$ is appropriately adjusted to raise or lower the temperature. Visual checks of the system are made using the program Visual Molecular Dynamics (VMD) created by the Theoretical and Computational Biophysics Group at the University of Illinois at Urbana-Champaign \cite{VMD}.

\begin{figure}[H]
	\centering
	\includegraphics[angle=0, scale=0.35]{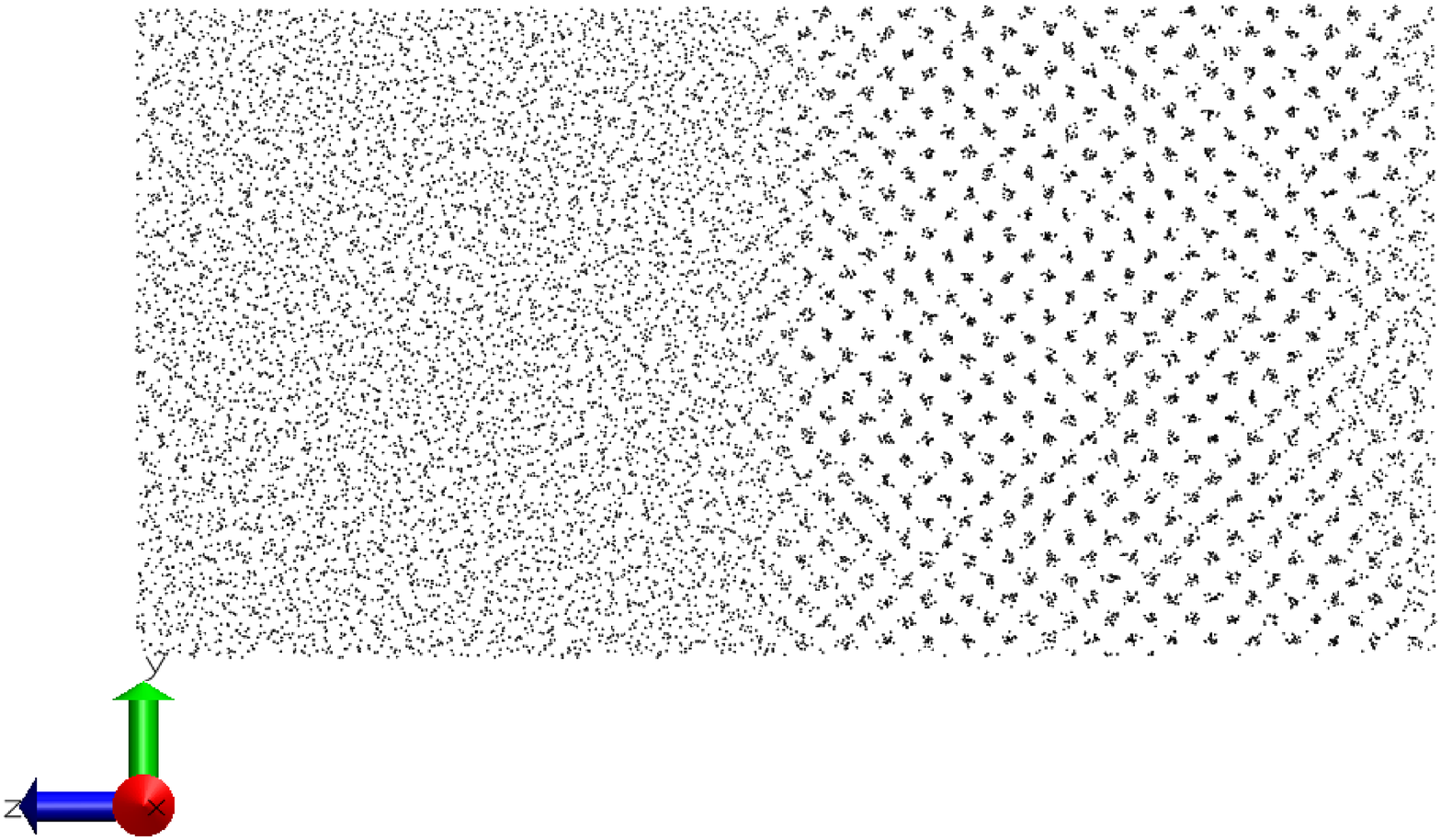}
	\vspace{-10pt}
\begin{singlespace}
	\caption[Two-Phase Configuration Visualized with VMD]{Sample configuration of N=16384 ions showing a liquid phase on the left and solid phase on the right. Figure prepared with VMD \cite{VMD}. \label{fig:VMD}}
\end{singlespace}
\end{figure}

\indent Initial conditions for both liquid and solid phases are created separately. Each phase consists of $8192$ ions for a total of $N=16384$ in the combined system. The liquid phase is created the same way as that described in section \ref{sec:liquiddiff}. The system is then cooled to $\Gamma=175$ and equilibrated for $1200/\omega_p$. The solid phase is acquired from the same initial conditions described in section \ref{sec:soliddiff} at $\Gamma=175$ as this configuration was readily available from previous simulations. The separate halves are brought together by adding a value of $L/2=8.92\lambda$ to the Z-component of the liquid configuration. Periodic boundary conditions along the Z-axis are altered to allow for a now rectangular box. A sample configuration of a two-phase system is given in Figure \ref{fig:VMD}.

\begin{figure}[H]
	\vspace{-15pt}
	\centering
	\includegraphics[angle=0, scale=0.5]{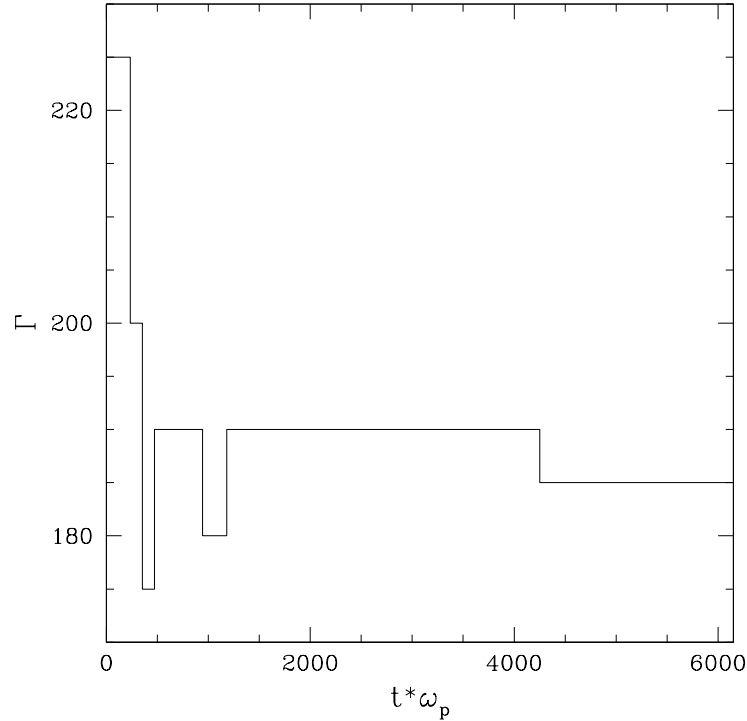}
	\vspace{-10pt}
\begin{singlespace}
	\caption[Melting Temperature Equilibration]{Range of temperatures used to bring the two-phase mixture of classically interacting ($\Lambda/a=0$) $^{16}$O ions near to its melting temperature. The system was given a total simulation time of $t\approx 6100/\omega_p$ (52,000 time steps). 	\label{fig:ClassicalEquil}}
\end{singlespace}
\end{figure}

\indent When the two halves are brought together the boundary between phases will have ions which are closer than would typically occur. This is because the two halves are equilibrated separately. Combined systems require time to equilibrate as the particles near boundaries adjust to their new neighbors. Shown in Figure \ref{fig:ClassicalEquil} is the range of temperatures and equilibration time given to bring the classically interacting two-phase simulation near the melting temperature.

\indent In order to understand how quantum corrections affect melting temperature, two-phase system with $\Lambda/a=0.335$ and $\Lambda/a=0.502$ were evolved. Initial conditions for simulations which include quantum corrections are created using the same method for initial conditions as the classical simulations described in the previous paragraph. Both liquid and solid phase configurations are equilibrated separately for $1200/\omega_p$ as before, except now, during the equilibration phase Eq. \eqref{eq2} is used for the inter-ion potential. The separate halves are also equilibrated at cooler temperatures than the classical simulation. A larger $\Gamma$ is chosen for this step because quantum corrections can melt the system due to zero point motion. This will be discussed in greater detail in chapters \ref{chap:results} and \ref{chap:discussion}. The two phases are then brought together as before and allowed to equilibrate while manually adjusting $\Gamma$ to keep the system half solid and half liquid.

\indent Both classical and semiclassical simulations are brought near to their melting temperatures by visually checking if the systems are changing phase and adjusting the value of $\Gamma$. Once the two-phase system is near the melting temperature it is then evolved microcanonically, at a constant energy, by no longer rescaling the ions' velocities. This method allows the release or absorption of latent heat allowing the system to adjust its own temperature and liquify or crystallize as needed. The interface between the liquid and solid, along with the temperature, will exponentially reach an equilibrium state \cite{Ercolessi}. Once the system has self-equilibrated, the melting temperature is inferred from the kinetic energy of the system. Results for these simulations are given in section \ref{melttempdis}.

%
	\chapter{Results}
\label{chap:results}
\graphicspath{{./Figures/}}

\indent \indent In this chapter we present results for all Molecular Dynamics (MD) simulations. Typical computation time for $N=8192$ ions and a simulation time of $t\approx 4800/\omega_p$ (40,000 steps) is approximately two-and-a-half days. Trajectory file sizes for liquid simulations are typically 8.3 GB in size. The authors wish to thank Indiana University's Scholarly Data Archive (SDA), formerly known as MDSS, for providing storage of data files. Calculation of the velocity autocorrelation function, which is complicated and very memory intensive, required on average thirteen hours. All simulations were performed on a Dell$^{\texttrademark}$ Optiplex 760 desktop. This 64-bit machine contained an Intel$\tiny{^{\textregistered}}$ Core$^{\texttrademark}$ 2 Duo CPU E8400 with 3.0 GHz processor speed along with 4.0 GB of RAM.

\section{\label{sec:V/N}Potential Energy per Particle}
\begin{figure}[H]
	\centering
	\includegraphics[angle=0, scale=0.45]{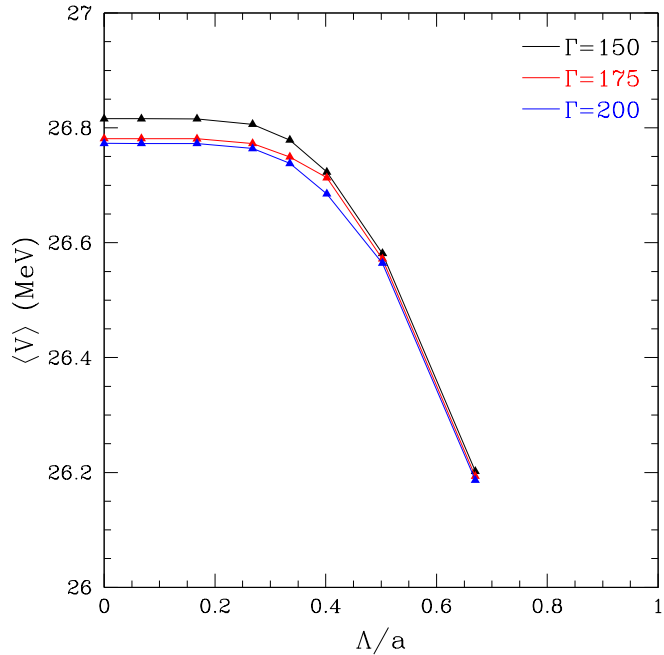}
	\vspace{-10pt}
\begin{singlespace}
	\caption[Potential Energy per Particle]{Potential energy per particle, $\langle V\rangle$, versus $\Lambda/a$ for all simulations. As quantum effects become larger the energy per particle continues to decrease due to decreased inter-ion potential. \label{fig:V/N}}
\end{singlespace}
\end{figure}

\indent \indent An important test for MD simulations is to evaluate the energy involved within a system. The average potential energy per particle $\langle V\rangle$ is calculated using
\begin{equation}
\langle V\rangle =\sum_{i<j}^{N} v_{ij}, \label{eq11}
\end{equation}
\noindent where $v_{ij}$ is equal to Eq. \eqref{eq1} for classical simulations or Eq. \eqref{eq2} for the semiclassical simulations. Figure \ref{fig:V/N}  and Table \ref{table1} show the change in $\langle V\rangle$ as a function of $\Lambda/a$ for the three temperatures used throughout this research. In crystals near the melting temperature thermal oscillations are large and ions come closer together and lead to higher total potentials. Thermal oscillations are smaller for lower temperatures--larger $\Gamma$--and ions stay farther apart. This decreases the total potential energy of the system. While the colder systems begin with a lower potential energy per particle, the characteristic to note is that the potential per particle decreases for increasing $\Lambda/a$. The average decrease in energy for all simulation temperatures is $0.596$ MeV. Both solid systems change phase due to zero point motion for large $\Lambda/a$, but the system's energy per particle changes very little. As the quantum influences become larger the potential becomes more attractive at short distances which decreases the potential per particle.

\vspace{-19pt}
\vspace{-25pt}
\begin{deluxetable}{ccccc}

\tablecolumns{4}
\tablewidth{0pc}
\tablecaption{\label{table1} $\langle V\rangle$ versus $\Lambda/a$ for All Systems}
\tablehead{
	\colhead{$\Lambda$ (fm)} & \colhead{$\Lambda/a$} & \colhead{$\Gamma =150$} & \colhead{$\Gamma =175$} & \colhead{$\Gamma =200$}\\
	\colhead{} & \colhead{} & \colhead{(MeV)} & \colhead{(MeV)} & \colhead{(MeV)}
}

\startdata
0 & 0 & 26.816 & 26.781 & 26.773\\
1 & 0.067 & 26.816 & 26.781 & 26.773\\
2.5 & 0.167 & 26.816 & 26.781 & 26.773\\
4 & 0.268 & 26.806 & 26.773 & 26.764\\
5 & 0.335 & 26.779 & 26.749 & 26.738\\
6 & 0.402 & 26.723 & 26.713 & 26.685\\
7.5 & 0.502 & 26.581 & 26.572 & 26.564\\
10 & 0.670 & 26.202 & 26.194 & 26.187\\
\enddata


\end{deluxetable}

\section{\label{liquidres}Diffusion in Liquids}
\begin{figure}[H]
	\centering
	\subfloat[Z(t) for several values of $\Lambda/a$.]{\label{fig:VACF}\includegraphics[width=0.49\textwidth]{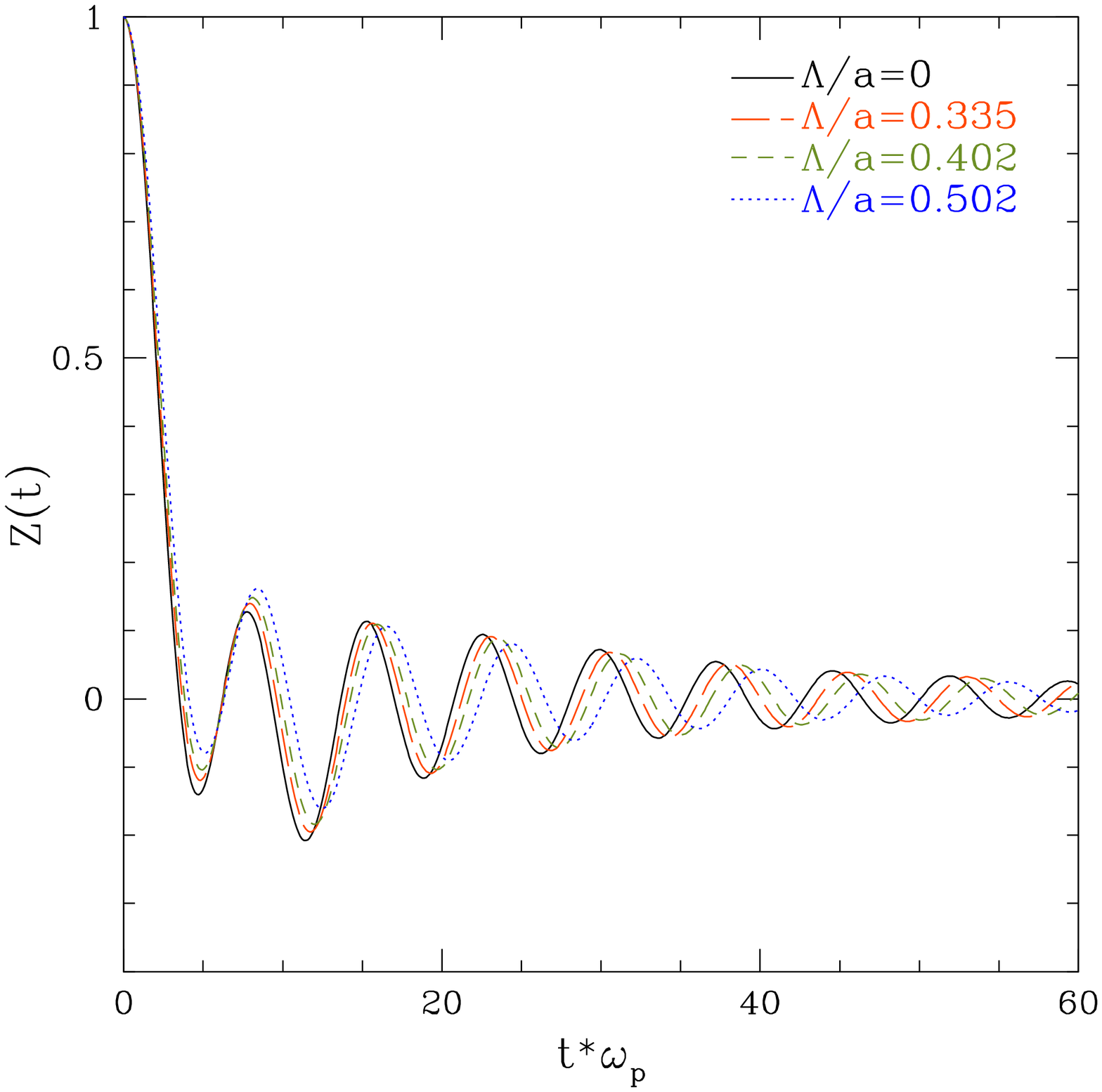}}
	\subfloat[Frequency of Z(t) scaled by $\omega_p$.]{\label{fig:w/wp}\includegraphics[width=0.49\textwidth]{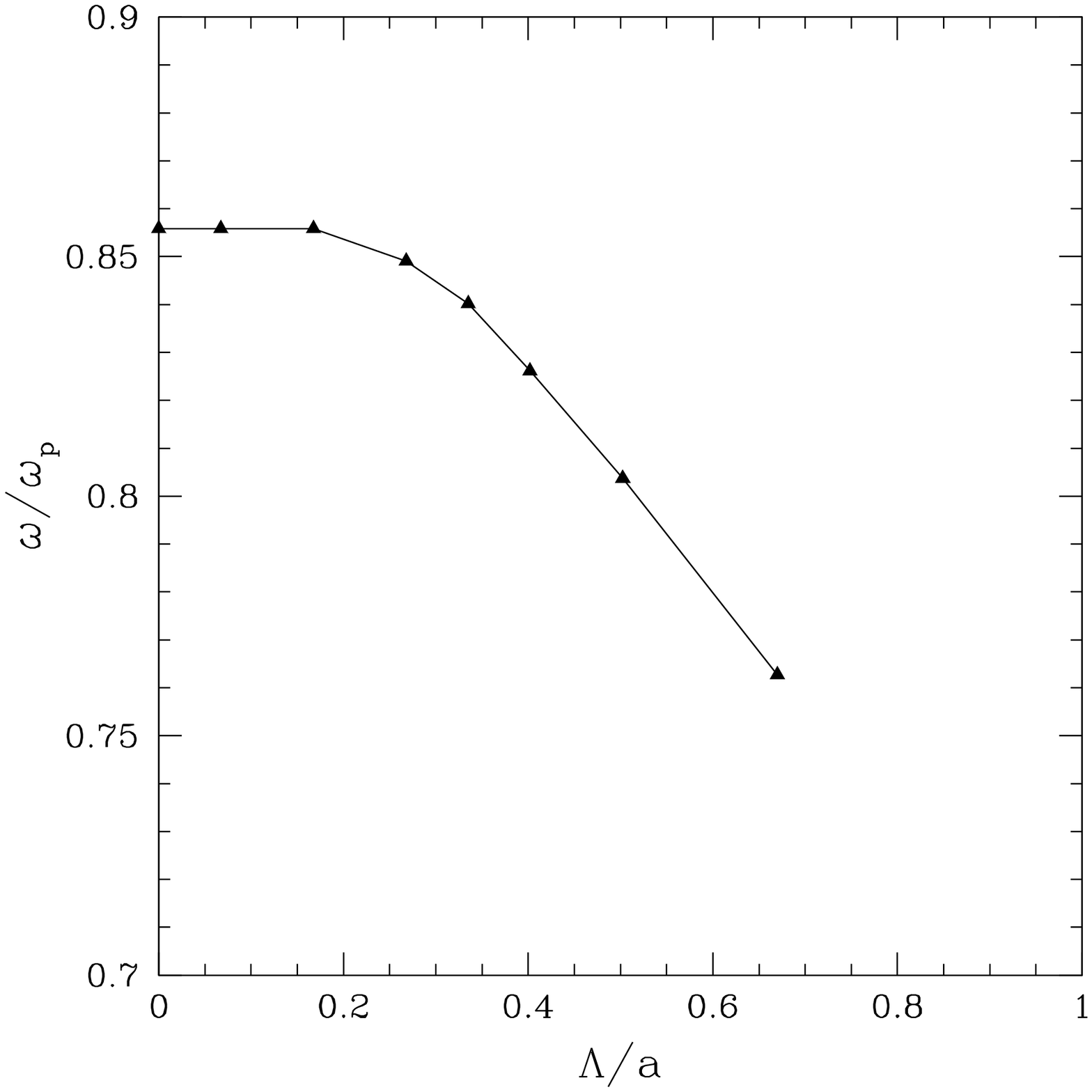}}
	\caption{Velocity Autocorrelation Function}
	\label{fig:VACF2}
\end{figure}

\indent \indent Figure \ref{fig:VACF} shows the velocity autocorrelation function $Z(t)$, Eq. \eqref{eq5}, for a classically interacting system and several intermediate values of $\Lambda/a$. Smaller values of $\Lambda/a$ are nearly indistinguishable from that of $\Lambda/a=0$ and are not shown. This figure shows that for increasing $\Lambda/a$, $Z(t)$ tends to `lag' behind that of a classically interacting system. The velocity autocorrelation function oscillates with a frequency near $\omega_p$, but zero point motion decreases the frequency with which $Z(t)$ oscillates. Figure \ref{fig:w/wp} shows how the frequency of the velocity autocorrelation function decreases as $\Lambda/a$ is increased. The plasma frequency, Eq. \eqref{eq4}, is used to scale all frequencies.
\begin{figure}[H]
	\centering
	\includegraphics[angle=0, scale=0.5]{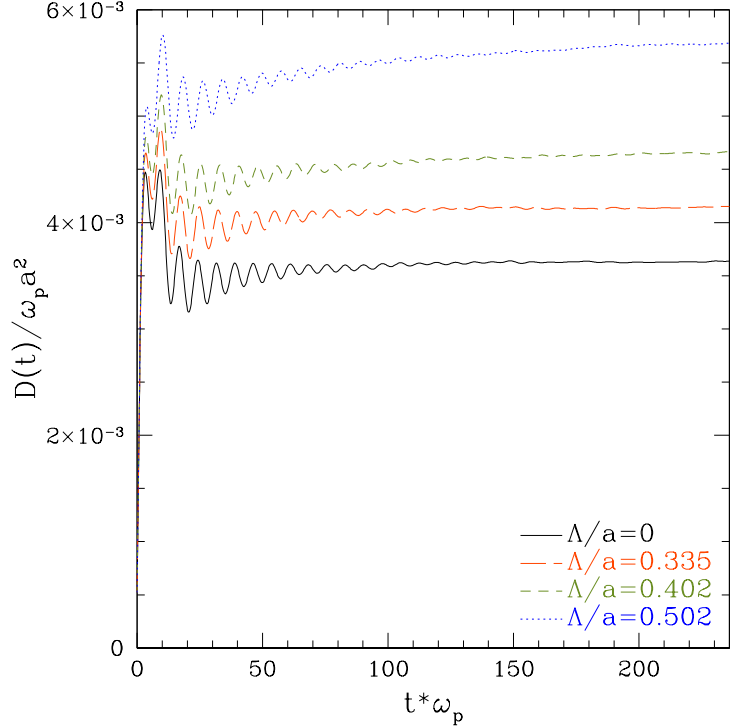}
	\vspace{-10pt}
\begin{singlespace}
	\caption[Diffusion in Liquid Phase Simulations]{Integral of Z(t),\eqref{eq6}, for several $\Gamma=150$ liquid simulations. Only values of $\Lambda/a$ which dramatically increased the diffusion are represented. \label{fig:LiquidDiffusion}}
\end{singlespace}
\end{figure}

\indent Figure \ref{fig:LiquidDiffusion} shows the integral of $Z(t)$, Eq. \eqref{eq6}. Values for $D$ are given in units of $\omega_pa^2$. Large fluctuations in $D(t)$ can be seen for small time $t$ due to large changes in $Z(t)$. As $t$ increases to $t_{max}$ $D(t)$ converges to that of the true value.

\begin{figure}[H]
	\centering
	\includegraphics[angle=0, scale=0.5]{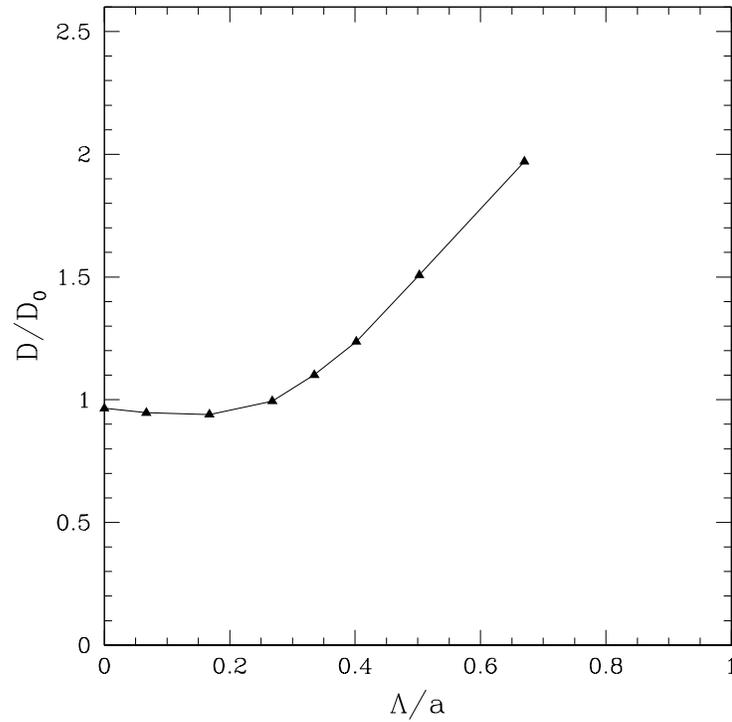}
	\vspace{-10pt}
	\caption[Liquid Diffusion Scaled by D$_0$]{Diffusion for liquid simulations scaled by D$_0$, Eq. \eqref{eq12}. \label{fig:D/D0}}
\end{figure}

\indent Diffusion constants for a one component plasma (OCP) in the liquid phase can be scaled by
\begin{equation}
D_0=\frac{3\omega_pa^2}{\Gamma^{4/3}}, \label{eq12}
\end{equation}
\noindent which is given by Hansen \emph{et al.}'s fit to their original MD results for diffusion \cite{Hansen 1975}. Figure \ref{fig:D/D0} and Table \ref{table2} show the scaled diffusion constants as a function of $\Lambda/a$. Diffusion in simulations with values of $\Lambda/a\lesssim 0.3$ deviates very little from Hansen \emph{et al.}'s classical, theoretical diffusion given by Eq. \eqref{eq12}. As quantum corrections increase, the diffusion slowly increases to twice the value predicted by Hansen \emph{et al.} when $\Lambda$ is two-thirds of the ion-sphere radius. The value of $\Lambda/a$ does not appear to be extremely important to the diffusion in a liquid system. The ions are allowed to move freely around one another in a liquid so the importance of zero point motion is reduced. Large values of $\Lambda/a$ help increase the diffusion, but the diffusion in a liquid is already relatively large.

\begin{deluxetable}{cccc}

\tablecolumns{4}
\tablewidth{0pc}
\tablecaption{\label{table2} Diffusion Constants for Liquid Phase Simulations}
\tablehead{
	\colhead{$\Lambda$ (fm)} & \colhead{$\Lambda$/a} & \colhead{$D/(\omega_pa^2)$} & \colhead{$D/D_0$}
}

\startdata
0 & 0 & 3.632 $\times\ 10^{-3}$ & 0.965 \\
1 & 0.067 & 3.564 $\times\ 10^{-3}$ & 0.947 \\
2.5 & 0.167 & 3.539 $\times\ 10^{-3}$ & 0.940 \\
4 & 0.268 & 3.742 $\times\ 10^{-3}$ & 0.994 \\
5 & 0.335 & 4.144 $\times\ 10^{-3}$ & 1.101 \\
6 & 0.402 & 4.654 $\times\ 10^{-3}$ & 1.237 \\
7.5 & 0.502 & 5.677 $\times\ 10^{-3}$ & 1.508 \\
10 & 0.670 & 7.417 $\times\ 10^{-3}$ & 1.970 \\
\enddata


\end{deluxetable}
\clearpage

\section{\label{sec:solidres}Diffusion in Solids}
\begin{figure}[H]
	\vspace{-10pt}
	\centering
	\includegraphics[angle=0, scale=0.5]{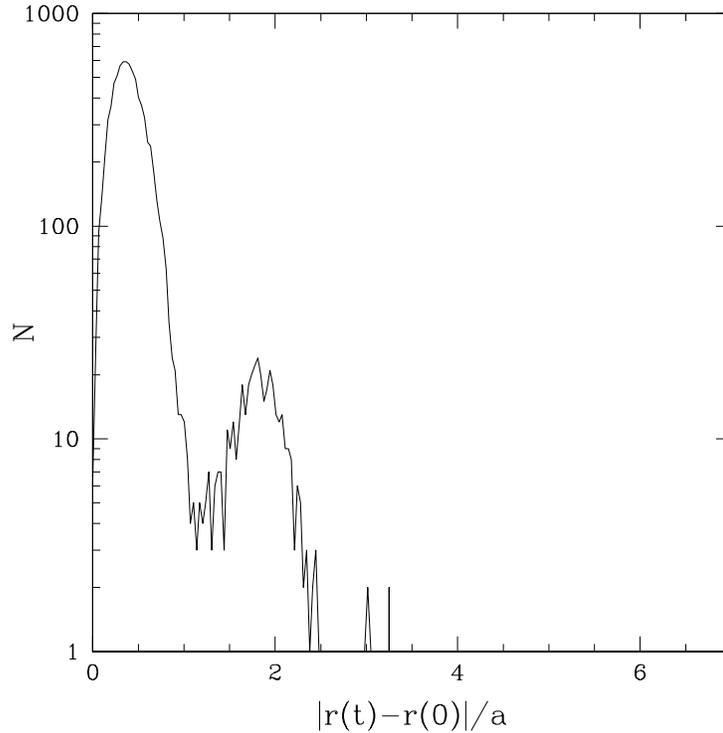}
	\vspace{-10pt}
\begin{singlespace}
	\caption[Histogram of Displacements for $\Lambda/a=0$]{Histogram of displacements $|r_j(t+t_0)-r_j(t_0)|$ in units of the ion-sphere radius $a$ for a classically interacting crystal lattice after a time $t\approx 4800/\omega_p$. \label{fig:Histogram0}}
\end{singlespace} 
\end{figure}

\indent \indent We begin by showing several histograms of displacements $|\textbf{r}_j(t+t_0)-\textbf{r}_j(t_0)|$ at the end of the simulations, $t\approx 4800/\omega_p$. These are computed by counting the number of ions that have moved a given distance in a time $t$. Figure \ref{fig:Histogram0} is the histogram for a purely classical system and shows a large central peak at small distances, which is due to each ion remaining at its original lattice site. The width of this peak corresponds to thermal oscillations of ions about their respective lattice sites. The smaller peak centered at $1.8a$ corresponds to ions that have ``hopped'' to neighboring lattice sites. A description of how ions move within a crystal lattice is given by Hughto \emph{et al.} (2011) \cite{Hughto 2011}. In this simulation a total of $372$ $\pm$19 ions ($\sim 4.5\%$ of the total number of ions) have moved farther than $1.07a$ from their original lattice site. This distance is also the cutoff distance used in Eq. \eqref{eq9}.

\begin{figure}[H]
	\centering
	\subfloat[Histogram of displacements for $\Lambda/a=0.335$.]{\label{fig:Histogram5}\includegraphics[width=0.49\textwidth]{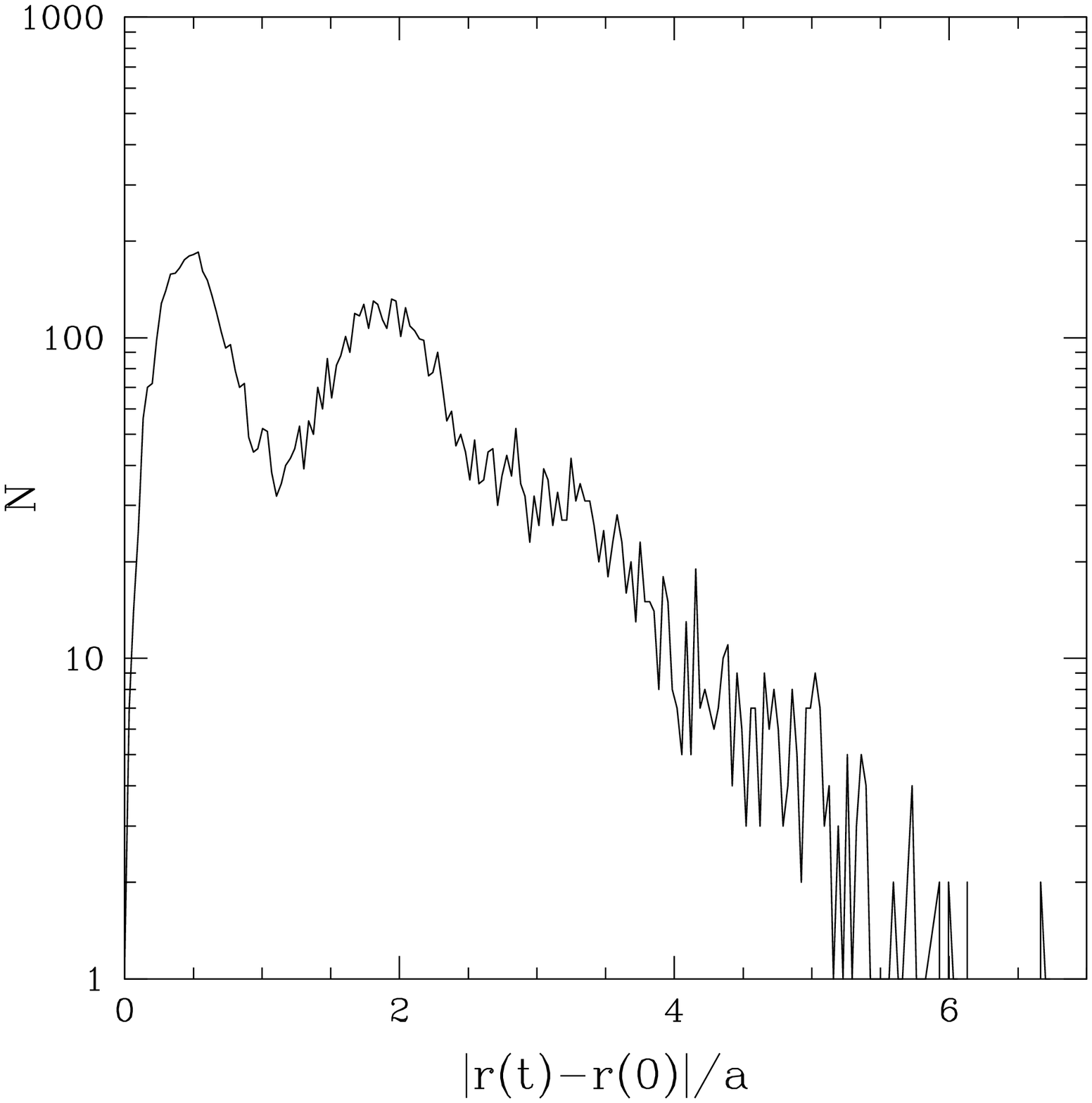}}
	\subfloat[Histogram of displacements for $\Lambda/a=0.402$.]{\label{fig:Histogram6}\includegraphics[width=0.49\textwidth]{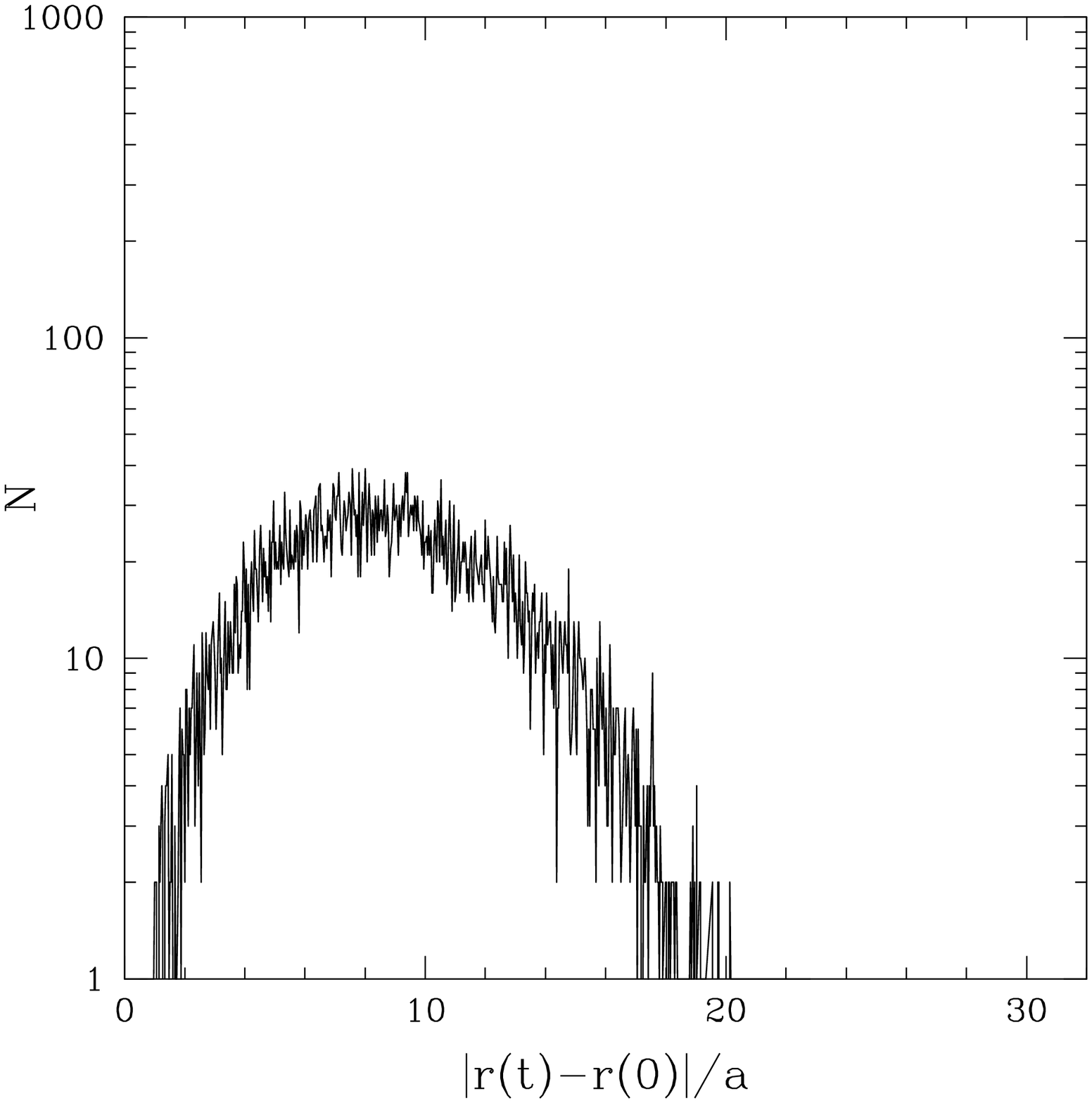}}
	\caption{Histogram of Displacements for the Semiclassical Potential}
	\label{fig:histograms}
\end{figure}

\indent Figure \ref{fig:Histogram5} shows the displacement histogram for $\Lambda/a=0.335$. The central peak for this histogram is smaller than the classical case, and the second peak is much larger than before. Many more ions are allowed to hop not just to neighboring lattice sites but to much farther distances. The distended peak out to $\sim 6a$ shows that many more ions have diffused in this semiclassical case. In this simulation $4405$ $\pm$66 ions ($\sim 54\%$) have diffused farther than the cutoff distance. As $\Lambda/a$ increases, the inter-ion potential decreases at intermediate distance and becomes more attractive at short distances allowing ions to diffuse more easily. As $\Lambda/a$ increases even further, the system can then be melted by zero point motion. This is shown in Figure \ref{fig:Histogram6} as there is only one extended peak centered around a distance of $\sim 8a$. At this point the zero point motion of ions has melted the crystal structure and ions can flow freely. Zero point motion is needed for ions to diffuse around neighboring particles, and large quantum corrections decrease the potential between ions, allowing diffusion to more readily occur. 

\begin{figure}[H]
	\vspace{-10pt}
	\centering
	\includegraphics[angle=0, scale=0.5]{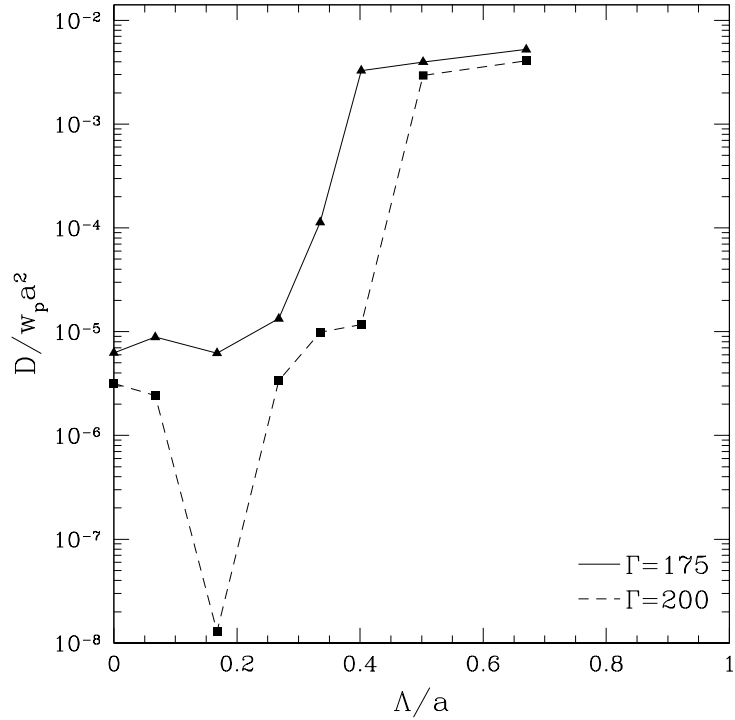}
	\vspace{-10pt}
	\caption[Diffusion in Solid Phase Simulations]{Diffusion constant $D$ versus $\Lambda/a$ for solid phase simulations. \label{fig:SolidDiffusion}}
\end{figure}

\indent Figure \ref{fig:SolidDiffusion} and Table \ref{table3} show the diffusion constants for all solid simulations. Diffusion in solids cannot be scaled by Eq. \eqref{eq12} because this equation only predicts diffusion in the liquid phase. For simulations with $\Gamma=175$ the diffusion changes very little while $\Lambda/a \lesssim 0.3$. At $\Lambda/a \approx 0.3$ the diffusion increases by a factor of 20 from a purely classical system, which can be seen as a large upturn in Figure \ref{fig:SolidDiffusion}. For $\Lambda/a \gtrsim 0.4$ the crystal structure is completely melted and the diffusion increases by a factor of 500. Once the system has melted, increasing $\Lambda/a$ further does not have a significant effect and the diffusion only increases slightly.

\indent Crystal simulations with $\Gamma=200$ exhibit similar characteristics. However, the system requires a larger value of $\Lambda/a$ before melting occurs. For a system to melt a combination of thermal and zero point motion is required. For the colder system the thermal motion is reduced; therefore, more quantum motion is required before melting can occur. This explains why $\Lambda/a$ must be larger to melt the colder system. We recognize the idea that the system may be super heating and address this later by calculating the melting temperature directly.

\indent Of particular note, the simulation with $\Gamma=200$ and $\Lambda/a=0.167$ exhibited very little diffusion. During the simulation several ions drifted farther than the cutoff distance and were considered to have diffused. Each of these ions subsequently returned to their original, vacant lattice sites within several more time steps. At the time of the simulation's end two ions had moved farther than the cutoff distance, giving a value for the diffusion that is two orders of magnitude smaller than any other simulation, $D/\omega_pa^2=1.6\times 10^{-8}$. It is likely that these two ions would have returned to their original lattice sites given more time.
\afterpage{
	\clearpage
		\begin{deluxetable}{cc|cc|ccc}

\tablecolumns{7}
\tablewidth{0pc}
\tablecaption{\label{table3} Diffusion Constants for Solid Phase Simulations}
\tablehead{
\colhead{} & \colhead{} & \multicolumn{2}{c}{$\Gamma=175$} & \colhead{} & \multicolumn{2}{c}{$\Gamma=200$} \\
\cline{3-4} \cline{6-7} \\
	\colhead{$\Lambda$ (fm)} & \colhead{$\Lambda$/a} & \colhead{$D/(\omega_pa^2)$} & \colhead{$\#$Ions Diffused} & &
	\colhead{$D/(\omega_pa^2)$} & \colhead{$\#$Ions Diffused}
}

\startdata
0 & 0 & 6.104 $\times\ 10^{-6}$ & 372 $\pm$19 && 3.171 $\times\ 10^{-6}$ & 196 $\pm$14\\
1 & 0.067 & 8.662 $\times\ 10^{-6}$ & 513 $\pm$23 && 2.421 $\times\ 10^{-6}$ & 150 $\pm$12\\
2.5 & 0.167 & 6.055 $\times\ 10^{-6}$ & 350 $\pm$19 && 1.584 $\times\ 10^{-8}$ & 2 $\pm$1\\
4 & 0.268 & 1.302 $\times\ 10^{-5}$ & 782 $\pm$28 && 3.377 $\times\ 10^{-6}$ & 199 $\pm$14\\
5 & 0.335 & 1.108 $\times\ 10^{-4}$ & 4405 $\pm$66 && 9.875 $\times\ 10^{-6}$ & 585 $\pm$24\\
6 & 0.402 & 3.193 $\times\ 10^{-3}$ & 8174 $\pm$90 && 1.175 $\times\ 10^{-5}$ & 691 $\pm$26\\
7.5 & 0.502 & 3.878 $\times\ 10^{-3}$ & 8172 $\pm$90 && 2.941 $\times\ 10^{-3}$ & 8170 $\pm$90\\
10 & 0.670 & 5.120 $\times\ 10^{-3}$ & 8183 $\pm$90 && 4.088 $\times\ 10^{-3}$ & 8179 $\pm$90\\
\enddata


\end{deluxetable}
	\clearpage
}
\clearpage

\section{\label{melttempres}Melting Temperature}
\begin{figure}[H]
	\centering
	\includegraphics[angle=0, scale=0.45]{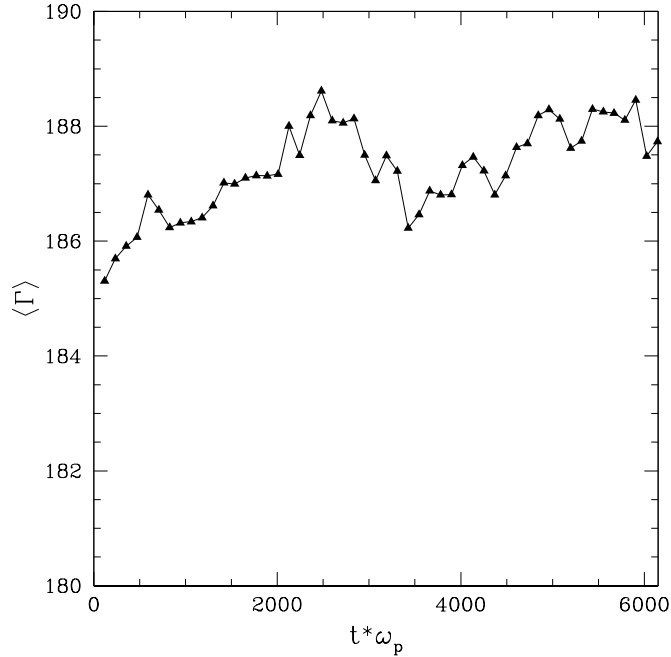}
	\vspace{-10pt}
\begin{singlespace}
	\caption[Classically Interacting Melting Temperature]{Average $\Gamma$ versus simulation time in a classical simulation. Each data point is the average of 100 values of $\Gamma$ spaced ten time steps apart. \label{fig:CEGamma0}}
\end{singlespace}
\end{figure}

\indent \indent Multi-phase systems are brought close to their equilibrium temperature by manually changing the value for $\Gamma$. Systems are then evolved at constant energy allowing the system to find an equilibrium temperature without external influences. When running simulations at constant energy a calculated value for $\Gamma$ is written to a file every 10 steps. Averages of $\Gamma$ are then made for every 1000 simulation steps and plotted versus time $t$. This shows how the system evolves and changes the temperature to reach an equilibrium point. Figure \ref{fig:CEGamma0} shows temperature changes throughout the simulation for a classical system while running at constant energy. Simulations are considered to have reached the melting temperature when changes in $\Gamma$ become small.
\vspace{-25pt}

\indent In our classically interacting system ($\Lambda/a=0$) the temperature initially behaves as expected, and increases towards the melting temperature. However, after a time $t=2100/\omega_p$ (18,000 time steps) the temperature begins to fluctuate. This system was given considerable time for fluctuations to diminish. To contest this the melting temperature is taken as the long-time average of many values. Only the final $t\approx 4000/\omega_p$ (34,000 time steps) are used in the average of $\Gamma$. Our inferred melting temperature of the classically interacting system is $\Gamma=187.6$ $\pm$1. The uncertainty is a conservative estimate that includes both the maximum and minimum of the fluctuations. This temperature is colder than the expected value of $\Gamma\approx 175$ \cite{Potekhin 2000}. Differences in our melting temperature from that of the expected value could be due to finite size effects and screening effects.

\begin{figure}[H]
	\centering
	\includegraphics[angle=0, scale=0.45]{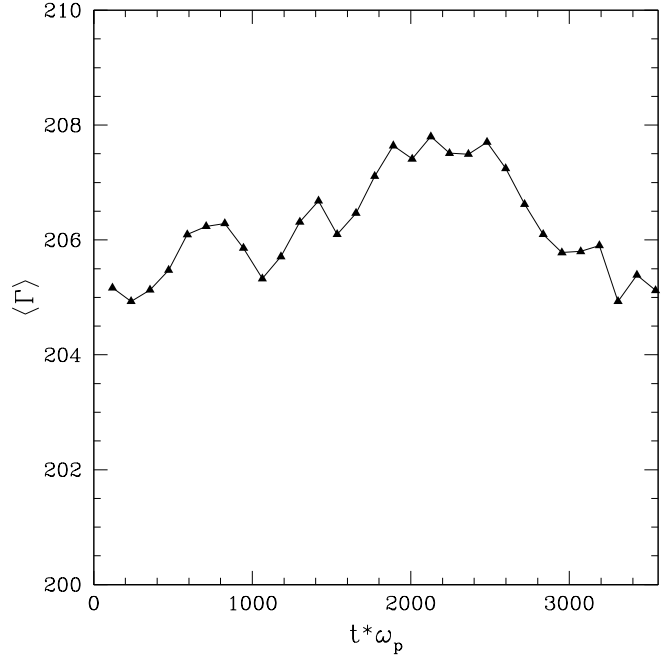}
	\vspace{-10pt}
\begin{singlespace}
	\caption[Melting Temperature for $\Lambda/a=0.335$]{Average $\Gamma$ versus simulation time in a semiclassical simulation with $\Lambda/a=0.335$. Each data point is the average of 100 values of $\Gamma$ spaced ten time steps apart. \label{fig:CEGamma5}}
\end{singlespace}
\end{figure}

\indent To understand how quantum corrections affect the melting temperature, a multi-phase simulation with $\Lambda/a=0.335$ has also been performed. This value was chosen because it is the stage in the solid simulations where diffusion was beginning to occur on a large scale basis. Figure \ref{fig:CEGamma5} shows the temperature of our $\Lambda/a=0.335$ simulation versus time. Fortunately, this simulation does not have the appreciable temperature fluctuations that the classical system shows in Figure \ref{fig:CEGamma0}, and, therefore, was run for a fraction of the computing time. From the $\Lambda/a=0.335$ simulation we infer a melting temperature of $\Gamma_m=206.5$ $\pm$1. A 10$\%$ increase in $\Gamma$ is needed to bring the system to an equilibrium condition. Again, the uncertainty is a conservative estimate which considers the maximum and minimum of the temperature fluctuations.

\begin{figure}[H]
	\centering
	\includegraphics[angle=0, scale=0.45]{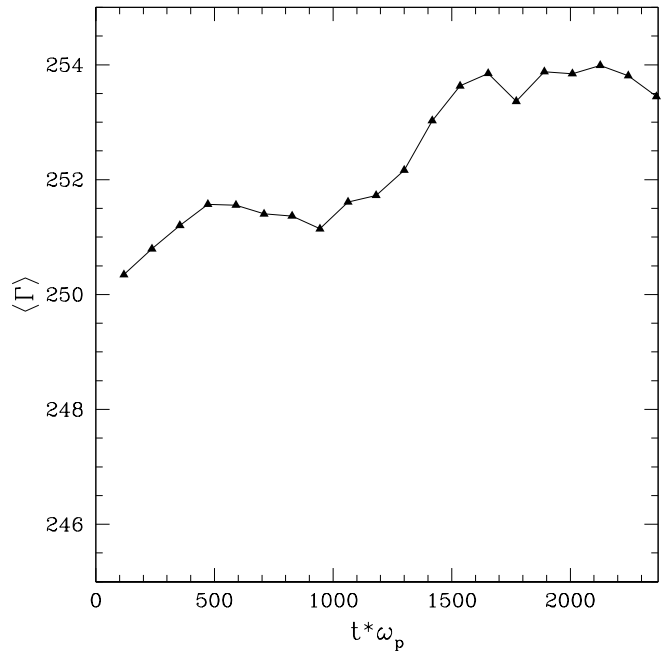}
	\vspace{-10pt}
\begin{singlespace}
	\caption[Melting Temperature for $\Lambda/a=0.502$]{Average $\Gamma$ versus simulation time in a semiclassical simulation with $\Lambda/a=0.502$. Each data point is the average of 100 values of $\Gamma$ spaced ten time steps apart. \label{fig:CEGamma7.5}}
\end{singlespace}
\end{figure}

\indent Finally, a multi-phase system with $\Lambda/a=0.502$ has also been run. This value was chosen because, at this stage, the zero point motion of the ions is large enough to melt both of our solid simulations. Shown in Figure \ref{fig:CEGamma7.5} is the temperature change of the $\Lambda/a=0.502$ simulation while evolving at constant energy. The melting temperature we infer from this data is $\Gamma_m=253.2$ $\pm$1. This corresponds to an increase of $35\%$ in the melting temperature. Table \ref{table5} gives the inferred melting temperatures versus quantum corrections. The implications of quantum corrections to the melting temperature are discussed further in chapter \ref{melttempdis}.
\begin{deluxetable}{ccc}

\tablecolumns{5}
\tablewidth{0pc}
\tablecaption{\label{table5}Melting Temperature versus Quantum Corrections}
\tablehead{
	\colhead{$\Lambda$ (fm)} & \colhead{$\Lambda$/a} & \colhead{$\Gamma_m$}
}

\startdata
0 & 0.0 & 187.6 $\pm$1\\
5 & 0.335 &  206.5 $\pm$1\\
7.5 & 0.502 & 253.2 $\pm$1\\
\enddata


\end{deluxetable}
%
	\chapter{Discussion}
\label{chap:discussion}

\section{\label{sec:Dependence} Quantum Dependence on $M$, $\Gamma$, and $n$}

\indent \indent In this section we go beyond the results of our Molecular Dynamics (MD) simulations and discuss the implications that this research has on our understanding of white dwarfs (WDs) and neutron stars (NSs). Recall that $\Lambda$=$\Lambda_{th}$/$\sqrt{2\pi^2}$ where $\Lambda_{th}$ is the ionic thermal deBroglie wavelength and is given by $\Lambda_{th}$= $\sqrt{2\pi\hbar^2/MT}$. Also recall that the ion-sphere radius is found by $a=(3/4\pi n)^{1/3}$. By combining these with a form of Eq. \eqref{eq3} that has been solved for $T$, we find 
\begin{equation}
\frac{\Lambda}{a}=\bigg[\frac{\hbar^2}{\pi} \frac{\Gamma}{MZ_iZ_je^2}\bigg(\frac{4\pi}{3}\bigg)^{1/3}\bigg]^{1/2}n^{1/6}, \label{eq13}
\end{equation}
\noindent where $M$ is the ionic mass, $Z_i$ the ionic charge, $\Gamma$ the Coulomb coupling parameter, and $n$ the ion density. From our simulations, specific values of $\Lambda/a$ have a respective associated diffusion, and in this chapter we consider several dependences of Eq. \eqref{eq13} to understand how each one can affect the diffusion of a system.

\subsection{\label{subsec:composition}Composition}
\begin{figure}[H]
	\centering
	\includegraphics[angle=0, scale=0.5]{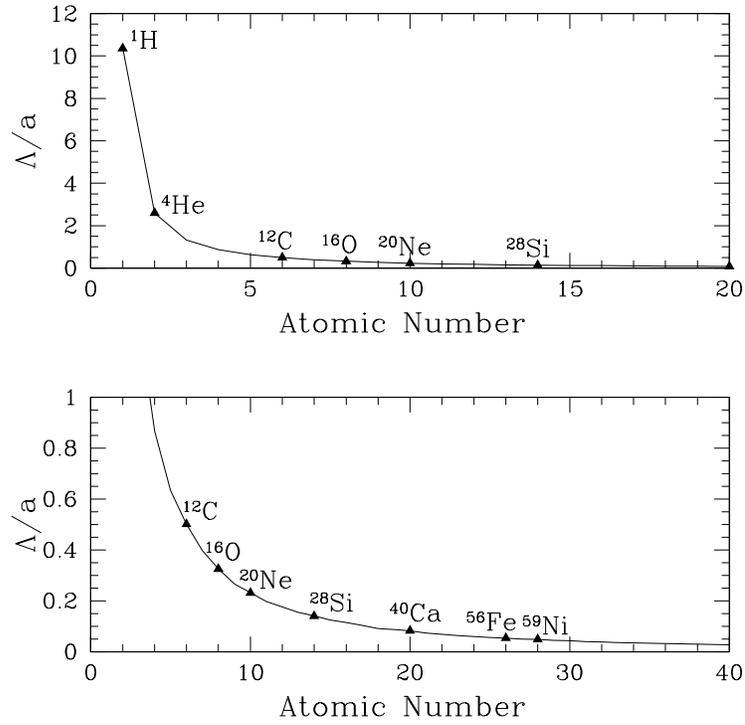}
	\vspace{-10pt}
\begin{singlespace}
	\caption[$\Lambda/a$ Dependence on Composition]{Values of $\Lambda/a$ versus atomic number (Z, M). The bottom graph contains the same data as the top graph, but it is scaled to the range $0\leq\Lambda/a \leq 1$. Calculations are performed at $\Gamma=175$ and an ion density $n=7.18\times 10^{-5}$ fm$^{-3}$. \label{fig:Composition}}
\end{singlespace}
\end{figure}

\indent \indent Here we discuss the $\Lambda/a$ dependence on the composition of a system. Various compositions will change $M$, $Z_i$ and $Z_j$ in Eq. \eqref{eq13}. This equation shows that $\Lambda/a$ $\propto$ $(Z_iZ_jM)^{-1/2}$. Therefore, heavier ions have smaller thermal deBroglie wavelengths compared to the ion-sphere radius. Calculations are done with $\Gamma=175$ and a relatively high ion density of $n=7.18\times 10^{-5}$ fm$^{-3}$. This density corresponds to a mass density of $1.91\times 10^{12}$ g cm$^{-3}$ and was chosen for historical reasons. Ionic mass $M$ is chosen using common stable isotopes of each element. Figure \ref{fig:Composition} shows how $\Lambda/a$ decreases with increasing atomic mass and ionic charge.

\indent It can be seen in Figure \ref{fig:Composition} that light elements can have large ionic thermal deBroglie wavelengths compared to the ion-sphere radius. The ion $^{4}$He can have a thermal deBroglie wavelength more than twice the ion-sphere radius and $^{1}$H can have a wavelength ten times the ion-sphere radius. These light elements are in the extreme quantum regime where small quantum corrections to the potential, such as that used in Eq. \eqref{eq2}, may be insufficient. On the other hand, heavy elements have smaller thermal deBroglie wavelengths. Ions such as $^{56}$Fe, at this density, have an associated $\Lambda/a \sim 0.05$ which puts it in the classical regime. It is the intermediate mass ions (eg. $^{12}$C and $^{16}$O) which fall into the semiclassical regime, where quantum corrections begin to become important, and with which this research is most concerned.

\subsection{\label{subsec:temperature}Temperature}
\begin{figure}[H]
	\centering
	\includegraphics[angle=0, scale=0.5]{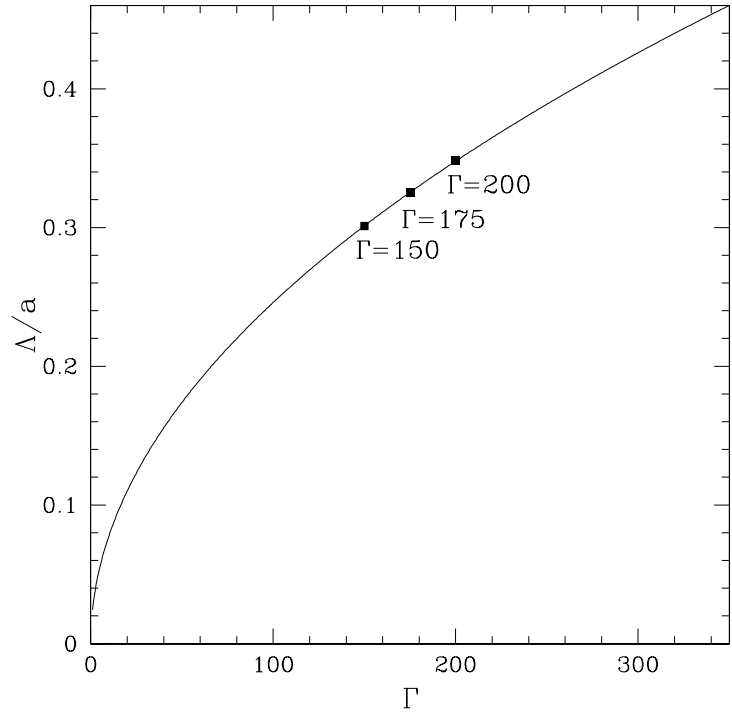}
	\vspace{-10pt}
\begin{singlespace}
	\caption[$\Lambda/a$ Dependence on Temperature]{Values of $\Lambda/a$ versus temperature $\Gamma$. Calculations performed with $^{16}$O at an ion density $n=7.18\times 10^{-5}$ fm$^{-3}$. \label{fig:TempDependence}}
\end{singlespace}
\end{figure}

\indent \indent Next we consider the dependence of $\Lambda/a$ on temperature, for which $\Gamma$ is strongly correlated through Eq. \eqref{eq3}. Increasing $\Gamma$ is analogous to decreasing the temperature and is very similar to a cooling WD. The density and composition of the system, aside from sedimentation, should change very little over time, but as it cools the value for $\Gamma$ will continue to increase. Eq. \eqref{eq13} shows that $\Lambda/a$ $\propto$ $\Gamma^{1/2}$. In calculations, we evaluate Eq. \eqref{eq13} using $^{16}$O and an ion density of $n=7.18\times 10^{-5}$ fm$^{-3}$. Figure \ref{fig:TempDependence} shows that at $\Gamma=150$ we find $\Lambda/a\approx 0.3$,  which is where quantum effects began increasing diffusion in our liquid simulations. This implies that a liquid may experience increasing diffusion as it cools and nears crystallization. This could also further chemical separation in the liquid crust of a NS \cite{Horowitz 2007,Medin 2010}. Furthermore, at $\Gamma=175$, $\Lambda/a$ is approximately $0.32$, which is where our simulation's diffusion had increased by a factor of 20. Finally, our $\Gamma=175$ simulations experience a complete phase transition at $\Lambda/a \gtrsim 0.4$. Eq. \eqref{eq13} does not evaluate to $0.4$ until $\Gamma\approx 265$. This semiclassical approximation implies that a WD could reliquify as it cools.

\subsection{\label{subsec:density}Density}
\indent \indent We now consider the $\Lambda/a$ dependence on ion density $n$. This is used to understand how $\Lambda/a$ behaves as you move to the interior of a WD or deeper into the crust of a NS. We first discuss the range of densities found in a WD or NS. WD stars typically have densities of $\rho\sim 10^6$ g cm$^{-3}$, but in an attempt to characterize ignition conditions of type Ia supernovae Lesaffre $\emph{et al.}$ \cite{Lesaffre 2006} find that for a range of stellar masses the central density for ignition for WDs falls within $2\times 10^{9}-5\times 10^{9}$ g cm$^{-3}$. A WD should not be expected to reach central densities above this or it would otherwise become a supernova \cite{Lesaffre 2006}. However, a NS is comprised of densities far beyond that of a WD. In a system where a NS is accreting mass from a companion star it will have a top liquid ocean layer; an outer crust which consists of a $^1$H burning layer at a density $\rho\sim 10^5$ g cm$^{-3}$; a level of $^4$He at $\rho\sim 10^6-10^8$ g cm$^{-3}$ which can burn unstably because it is strongly degenerate; and a bottom layer with $\rho\sim 10^{11}$ g cm$^{-3}$ where neutronization becomes relevant, ions are neutron rich, and neutron drip begins \cite{Zdunik 2008}. The inner crust densities reach $\rho\sim 10^{13}-10^{14}$ g cm$^{-3}$ and become a homogeneous mixture  of $n$, $p$, and $e^-$ with few percent protons at the transition to the core \cite{Zdunik 1990}. The bottom of the inner crust is near normal nuclear density $\rho =2.7\times 10^{14}$ g cm$^{-3}$ \cite{Zdunik 2008}.

\indent It is also important to mention the densities at which pycnonuclear reactions occur ($\rho_{pyc}$) and densities for neutronization ($\rho_n$). Pycnonuclear reactions are heavily dependent on density rather than temperature. Thermal vibrations of ions about their lattice, along with the probability to tunnel through the repulsive Coulomb barrier, can lead to nuclear reactions. Reactions set in quickly at $\rho > \rho_{pyc}$. Typical densities for pycnonuclear reactions are $\rho_{pyc}\approx 10^6, 10^9$, and $10^{10}$ g cm$^{-3}$ for burning $^1$H, $^4$He, and $^{12}$C respectively. Because of their larger Coulomb barriers heavy elements require higher $\rho_{pyc}$. For each ion there is also a threshold $\rho_n$ of the density for which neutronization occurs. For $^1$H, $^4$He, and $^{12}$C this density is $\rho_n\approx 1.2\times 10^7$, $1.4\times 10^{11}$, and $3.9\times 10^{10}$ g cm$^{-3}$ respectively. For all ions mentioned, $\rho_{pyc}\ll \rho_n$ and pycnonuclear reactions occur before neutronization \cite{Kippenhahn 1994}.

\indent The thermal deBroglie wavelength, $\Lambda_{th}=\sqrt{2\pi\hbar^2/MT}$, does not explicitly depend on the density of the system. However, the temperature $T$ does have a density dependence. Eq. \eqref{eq13} shows that $\Lambda/a$ $\propto$ $n^{1/6}$. The size of the thermal deBroglie wavelength is decreased for dense systems, but not as rapidly as the ion-sphere radius. For large densities the ratio $\Lambda/a$ continues to increase and quantum corrections become more important.

\begin{figure}[H]
	\centering
	\includegraphics[angle=0, scale=0.5]{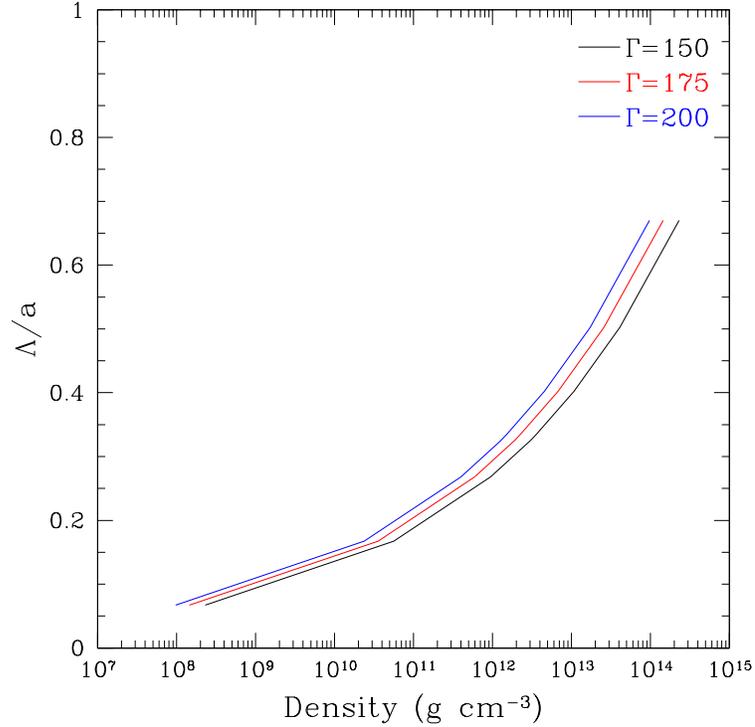}
	\vspace{-10pt}
\begin{singlespace}
	\caption[$\Lambda/a$ Dependence on Density]{Values of $\Lambda/a$ versus density $\rho$. Calculations are performed with $^{16}$O. \label{fig:DensityDependence}}
\end{singlespace}
\end{figure}

\indent To get an initial perspective on what density range quantum effects become important to diffusion we use Eq. \eqref{eq13} and solve for the ion density $n$. For a given ion, $^{16}$O, at several $\Gamma$, the density dependence of $\Lambda/a$ is plotted in Figure \ref{fig:DensityDependence}. This figure shows a wide range of densities needed for $\Lambda/a$ to increase from the classical regime, through the semiclassical, and into the quantum regime. An interesting thing to note in Figure \ref{fig:DensityDependence} is that for liquids, $\Gamma =150$, higher densities are consistently needed to achieve the same $\Lambda/a$ as crystal structures because of the higher temperatures.

\begin{figure}[H]
	\centering
	\includegraphics[angle=0, scale=0.5]{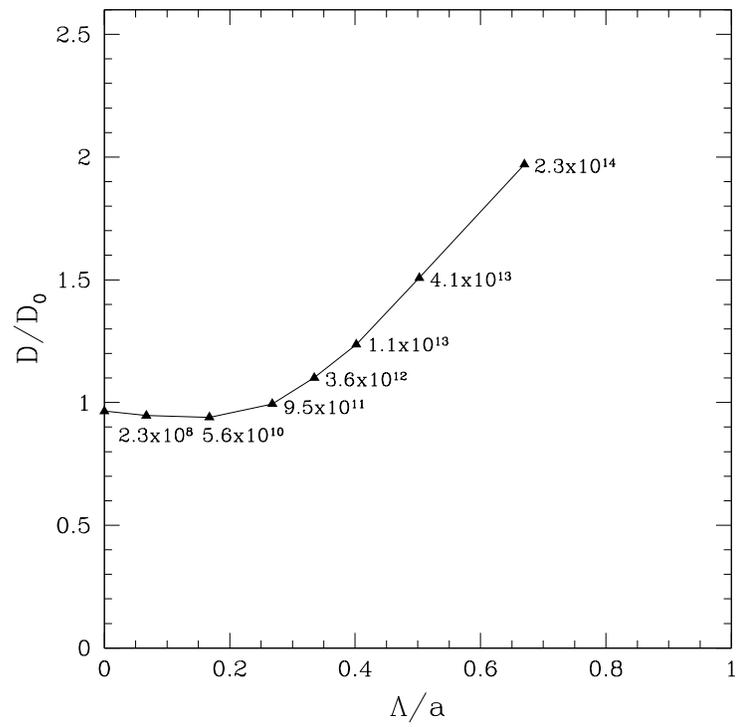}
	\vspace{-10pt}
\begin{singlespace}
	\caption[Diffusion in Liquid Phase Simulations with Calculated Densities]{Diffusion constants for liquid phase simulations with densities calculated from Eq. \eqref{eq13}. Calculations performed with $^{16}$O and $\Gamma=150$. All densities are given in units of g cm$^{-3}$. \label{fig:LiquidDensity}}
\end{singlespace}
\end{figure}

\begin{figure}[H]
	\centering
	\includegraphics[angle=0, scale=0.5]{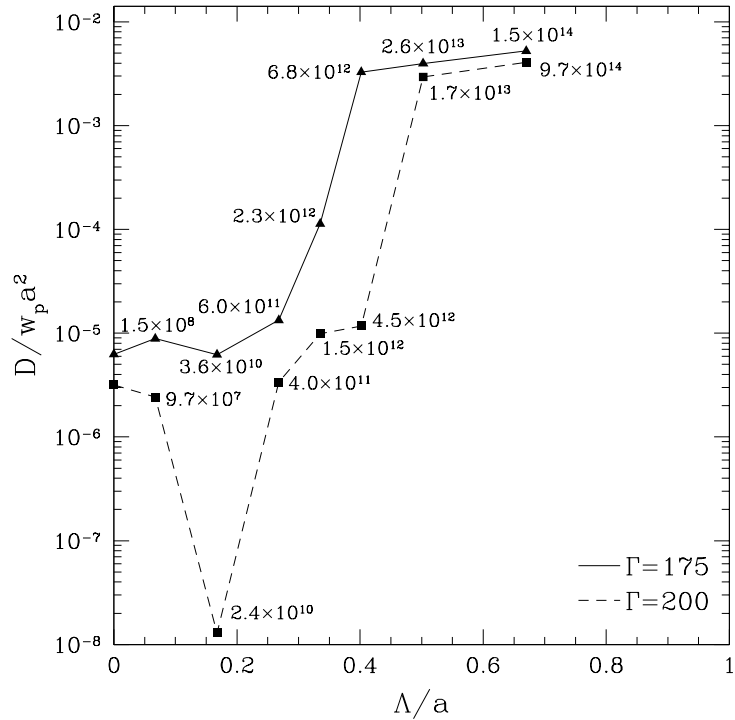}
	\vspace{-10pt}
\begin{singlespace}
	\caption[Diffusion in Solid Phase Simulations with Calculated Densities]{Diffusion constants for solid phase simulations with densities calculated from Eq. \eqref{eq13}. Calculations performed with $^{16}$O, $\Gamma=175$, and $\Gamma=200$. All densities are given in units of g cm$^{-3}$. \label{fig:SolidDensity}}
\end{singlespace}
\end{figure}

\indent From our simulations, we can go further and now associate diffusion with density. Figures \ref{fig:LiquidDensity} and \ref{fig:SolidDensity} are the same as Figures \ref{fig:D/D0} and \ref{fig:SolidDiffusion} respectively and show diffusion versus $\Lambda/a$ for all simulation temperatures. However, Eq. \eqref{eq13} has been used to calculate the density at each value of $\Lambda/a$, at its respective $\Gamma$. These densities are also given in Table \ref{table3}. All simulations were composed of a OCP of $^{16}$O, therefore, all calculations for densities in these figures and table use a OCP of $^{16}$O as well.

\begin{deluxetable}{ccccc}

\tablecolumns{5}
\tablewidth{0pc}
\tablecaption{\label{table4} Calculated Densities from Eq. \eqref{eq13}.}
\tablehead{
	\colhead{$\Lambda$ (fm)} & \colhead{$\Lambda$/a} & \colhead{$\Gamma=150$} & \colhead{$\Gamma=175$} & \colhead{$\Gamma=200$}
}

\startdata
1 & 0.067 & 2.3 $\times$ 10$^{8}$ & 1.5 $\times$ 10$^{8}$ & 9.7 $\times$ 10$^{7}$\\
2.5 & 0.167 & 5.6 $\times$ 10$^{10}$ & 3.6 $\times$ 10$^{10}$ & 2.4 $\times$ 10$^{10}$\\
4 & 0.268 & 9.5 $\times$ 10$^{11}$ & 6.0 $\times$ 10$^{11}$ & 4.0 $\times$ 10$^{11}$\\
5 & 0.335 & 3.6 $\times$ 10$^{12}$ & 2.3 $\times$ 10$^{12}$ & 1.5 $\times$ 10$^{12}$\\
6 & 0.402 & 1.1 $\times$ 10$^{13}$ & 6.8 $\times$ 10$^{12}$ & 4.5 $\times$ 10$^{12}$\\
7.5 & 0.502 & 4.1 $\times$ 10$^{13}$& 2.6 $\times$ 10$^{13}$ & 1.7 $\times$ 10$^{13}$\\
10 & 0.670 & 1.5 $\times$ 10$^{14}$ & 1.5 $\times$ 10$^{14}$ & 9.8 $\times$ 10$^{13}$\\
\enddata


\end{deluxetable}

\indent Finally we describe the diffusion dependence not only on density but on both density and composition. It has already been shown that light elements can have relatively large thermal deBroglie wavelengths, and that as the density of a system increases quantum corrections become more important allowing diffusion to occur more readily. We do this to understand what densities various ions require before quantum effects become important. This work is complementary since sedimentation in WDs \cite{Winget 2009} and phase separation in the crust of a NS \cite{Horowitz 2007} are expected to occur for different compositions.

\begin{figure}[H]
	\centering
	\includegraphics[angle=0, scale=0.6]{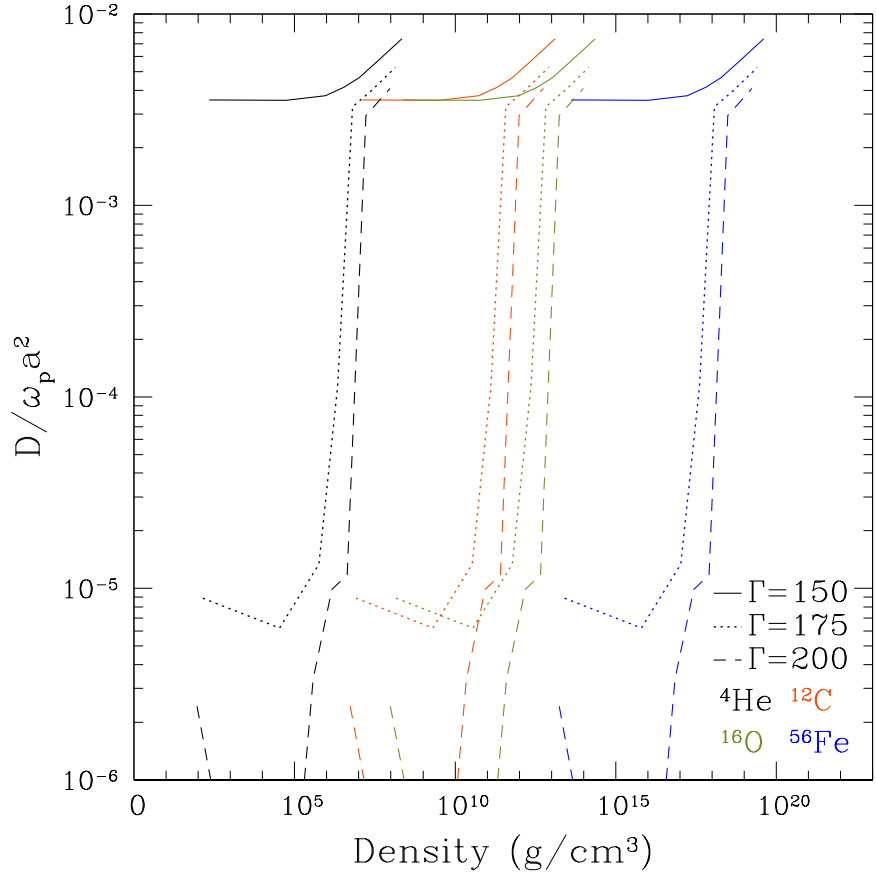}
	\vspace{-10pt}
\begin{singlespace}
	\caption[Diffusion Constants for All Simulations as a Function of Density]{Diffusion constants for all simulations as a function of density. The type of line (\emph{i.e.} solid versus dashed) represents the temperature of the simulation. The color of the line represents the ion (Z, M) used in the density calculations. \label{fig:CompDensity}}
\end{singlespace}
\end{figure}

This is done by choosing a specific ion (Z, M), holding $\Gamma$ constant, holding $\Lambda/a$ constant, and then solving for the density $n$ in Eq. \eqref{eq13}. Figure \ref{fig:CompDensity} shows the diffusion dependence on density for common elements within a WD or NS crust (\emph{eg}. $^{4}$He, $^{12}$C, $^{16}$O, and $^{56}$Fe).

\indent It can be seen that light ions require low densities before quantum corrections become important. For example, $^{4}$He at a $\Gamma=175$ requires a density $\rho\sim 2\times 10^6$ g cm$^{-3}$ before zero point motion can melt the system. This density occurs in both WDs and NS crusts and is below the threshold densities $\rho_{pyc}$ and $\rho_n$.

\indent Figure \ref{fig:CompDensity} shows that $^{56}$Fe requires $\rho\sim 10^{17}$ g cm$^{-3}$ for all simulation temperatures before zero point motion becomes important. However, $^{56}$Fe has a density for neutronization ($\rho_n\approx 1.14\times 10^9$ g cm$^{-3}$) that is below its pycnonuclear reaction density  \cite{Kippenhahn 1994}. Therefore, neutronization of $^{56}$Fe occurs before nuclear reactions occur. The density needed for quantum corrections to become important is far beyond the neutronization density and the density achieved by a NS.

\indent Intermediate mass ions (\emph{i.e.} $^{12}$C and $^{16}$O) need much lower densities than $^{56}$Fe before quantum corrections become important. A solid system composed mostly of $^{12}$C near the melting temperature, $\Gamma=175$, could remain a liquid if the density is over $\sim 1\times 10^{11}$ g cm$^{-3}$. Such a density is beyond that of a WD, but it can be found in the bottom of the outer crust of a NS. However, this density is greater than $\rho_{pyc}$ for $^{12}$C and nuclear reactions should occur first. We find that for both $^{12}$C and $^{16}$O the densities required for relevant quantum corrections are beyond that of the WD peak central densities stated by Lesaffre \emph{et. al} \cite{Lesaffre 2006}. However, zero point motion is relevant in the bottom of the outer crust of a NS where such densities can be found.

\section{\label{melttempdis}Quantum Effects on Melting Temperature}
\begin{figure}[H]
	\centering
	\includegraphics[angle=0, scale=0.49]{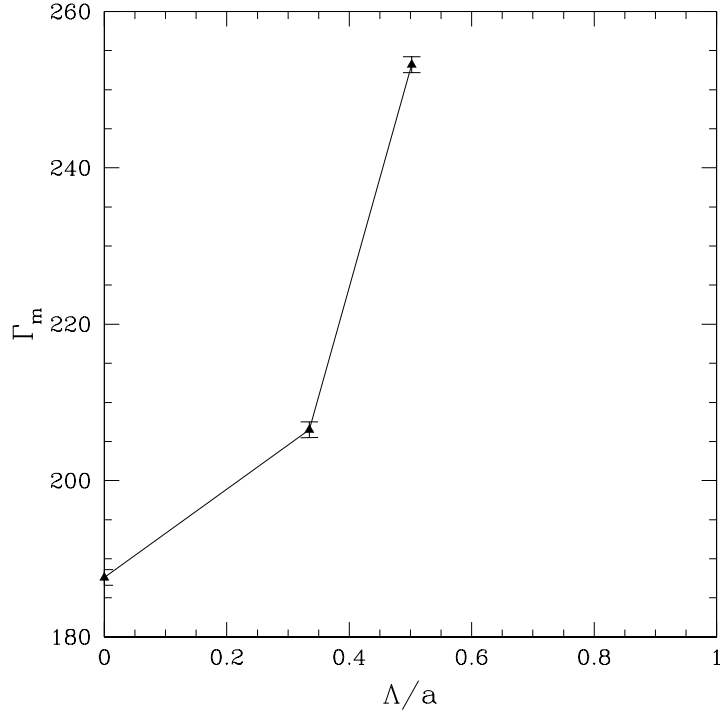}
	\vspace{-10pt}
\begin{singlespace}
	\caption[Quantum Corrections to the Melting Temperature]{Melting Coulomb parameter $\Gamma_m$ for two-phase simulations. All simulations contain $N=16384$ ions. Error bars are $\pm$1 in magnitude and are a conservative estimate on the statistical error on $\Gamma_m$. \label{fig:MeltingTemp}}
\end{singlespace}
\end{figure}

\indent\indent We now discuss how quantum corrections influence the melting temperature. Figure \ref{fig:MeltingTemp} depicts the melting Coulomb parameter $\Gamma_m$ of our two-phase simulations versus $\Lambda/a$. 

\noindent We find that $\Gamma_m$ increases rapidly for $\Lambda/a>0.3$. This trend in Figure \ref{fig:MeltingTemp} may be much smoother than represented if simulations were to be run at more values of $\Lambda/a$. Modifying the melting temperature of material in a WD could play an important role in its evolution because the cooling mechanism depends on the phase of matter. This alters when the radiation of the latent heat of fusion will delay cooling \cite{Salpeter 1961}. However, quantum corrections do not impact melting temperature significantly until $\Lambda/a\sim 0.335$ which corresponds to a density of $2.3\times 10^{12}$ g cm$^{-3}$for $^{16}$O at $\Gamma=175$. This is higher than the densities reached in a WD, and, therefore, the ions in C and O WDs can be considered classical. For light elements, such as $^4$He, the density required is only of order 10$^6$ g cm$^{-3}$, and melting temperatures could be altered by quantum effects. This implies that a He white dwarf would not freeze, and there would be no delay in the cooling curve as it radiates latent heat.

\indent We now compare our melting temperature results to that of Jones and Ceperley (1996) \cite{Jones 1996}. Their work used path integral Monte Carlo (PIMC) simulations to study the OCP at finite temperature and directly calculate the importance of quantum effects on melting temperatures in two-phase systems. Simulations are described by the dimensionless ratio $r_s=a/a_0$, where $a$ is the ion-sphere radius and $a_0$ is a natural length scale given by $a_0=\hbar^2/mZ^2e^2$. Large values of $r_s$ correspond to more classical systems while smaller values are more quantum in nature. In Jones and Ceperley's Fig. 1, they find that the melting temperature of a OCP is similar to the classical prediction for a large portion of their phase diagram. This figure also shows that for a given temperature as the density increases a OCP will solidify. Increasing the density even further increases quantum effects and can reliquify the system. This is in agreement with our results in chapter \ref{subsec:density}. At very high densities $\Lambda/a$ becomes large and the melting temperature becomes very low, $\Gamma_m$ approaches infinity. Even at $T=0$ the system can remain a quantum fluid.

\indent Given in Table \ref{table6} is a comparison of $r_s$ to $\Lambda/a$ for all simulation temperatures. Note that large values of $\Lambda/a$ correspond to small values of $r_s$. However, from our simulations when $\Lambda/a=0.335$ we find a 10$\%$ increase in $\Gamma_m$ from the classical case, see Table \ref{table5}. At the corresponding density from chapter \ref{subsec:density} this gives a value of $r_s\sim 590$. Jones and Ceperley do not find a noticeable change in $\Gamma_m$ until somewhat smaller values, $r_s\sim 200$. In the case of $\Lambda/a=0.502$ we find a $35\%$ increase in $\Gamma_m$. For this condition's corresponding density we find $r_s\sim 320$. Comparing our data to that of Fig. 1 in Jones and Ceperley our predictions for the melting temperature fall between the semiclassical prediction of ref. \cite{Chabrier 1993} and that of the full quantum calculations.

\begin{deluxetable}{ccccc}

\tablecolumns{5}
\tablewidth{0pc}
\tablecaption{\label{table6} Values of $r_s$ Corresponding to $\Lambda/a$}
\tablehead{
	\colhead{$\Lambda$ (fm)} & \colhead{$\Lambda$/a} & \colhead{$\Gamma=175$} & \colhead{$\Gamma=206.5$} & \colhead{$\Gamma=251.5$}
}

\startdata
1 & 0.067 & 12400 & 14640 & 17830\\
2.5 & 0.167 & 1990 & 2340 & 2850\\
4 & 0.268 & 780 & 920 & 1110\\
5 & 0.335 & 500 & 590 & 710\\
6 & 0.402 & 350 & 410 & 500\\
7.5 & 0.502 & 220 & 260 & 320\\
10 & 0.670 & 120 & 150 & 180\\
\enddata


\end{deluxetable}

%
	\chapter{Summary and Conclusions}
\label{chap:conclusion}

\indent\indent Quantum corrections to the inter-ion potential can be important for dense astrophysical objects and can affect the amount of diffusion in these systems. We have performed MD simulations of a OCP to understand the impact of quantum corrections. Diffusion coefficients for liquid and solid phase simulations have been calculated. Quantum corrections in the liquid configurations do not dramatically affect diffusion. Quantum corrections depend on the parameter $\Lambda$ which is related to the ionic thermal deBroglie wavelength of an ion. Simulations show that the diffusion coefficient only increases by a factor of two for $\Lambda/a=0.670$. However, quantum corrections in solid systems are much more important. We find that for systems near the melting temperature increasing $\Lambda/a$ to $0.335$ increases the diffusion coefficient by a factor of 20, and for $\Lambda/a=0.402$ the system can be melted.

\indent Shortly after crystallization in the core of WDs, where $\Gamma\sim 200$, quantum effects are unimportant unless $\rho\gtrsim 1.5\times 10^{12}$ g cm$^{-3}$ for $^{16}$O, $1.5\times 10^{6}$ for $^{4}$He, and $2.7\times 10^{17}$ for $^{56}$Fe. Quantum effects for ions between $^{16}$O and $^{56}$Fe are likely to be small at WD densities. Quantum corrections for $^{4}$He could be larger if $^{4}$He survives to high densities. In NSs, quantum effects for $^{16}$O should be considered as the essential density can occur in the inner crust. However, the density needed for zero point motion to be significant for $^{56}$Fe is well beyond that reached by WDs or NSs.

\indent It is important to understand the role of quantum corrections to the melting temperature in dense systems. Altering the melting temperature of a WD determines how long the star has to cool before it crystallizes, and in the crust of a NS it can affect the structure and depth at which crystallization occurs. We determined melting temperatures by performing two phase MD simulations where both liquid and solid phases are equilibrated simultaneously. We find that increasing the quantum corrections to $v_{ij}$ increases $\Gamma_m$. For $\Lambda/a=0.335$ we find $\Gamma_m=206.5$ $\pm$1, a $10\%$ increase, and for $\Lambda/a=0.502$ we find $\Gamma_m=250$ $\pm$1, a $35\%$ increase. For $^{16}$O this corresponds to densities of $1.4\times 10^{12}$ and $8.7\times 10^{12}$ g cm$^{-3}$ respectively, which can be found in NSs. For larger $\Lambda/a$ the value of $\Gamma_m$ may become very large. Indeed, at very high density there can be a quantum fluid that remains liquid even at zero temperature.

\indent In conclusion, we find quantum corrections to be small in WDs unless considering light elements such as $^4$He. The quantum effects in the crust of NSs should be considered for light-to-intermediate elements (e.g. $^{12}$C and $^{16}$O), if these ions survive to high densities in the inner crust. For ions heavier than $^{56}$Fe the densities required for quantum corrections appear to be higher than achieved by a NS and the ions can be considered classical.

	\begin{vitae}
		\section*{\sc Contact Information}
		\begin{tabular}{@{}p{2.8in}p{2.8in}}
			3166 E. Covenanter Dr. & {\it Phone:} +1 (812) 738-9438 \\
			Bloomington, IN 47401  & {\it E-mail:}  jrmason@indiana.edu \\
		\end{tabular}

		\section*{\sc Education}
		{\bf Indiana University}, Bloomington, IN, USA \\
		\begin{cvlist}
			\item[] M.A. Astronomy, August 2011
			\begin{cvlistb}
				\item Thesis Title:  ``Quantum Corrections to Diffusion in Stars'' 
				\item Advisor:  Dr. Charles Horowitz, Director Nuclear Theory Center
			\end{cvlistb}
		\end{cvlist}
		
		{\bf Ball State University}, Muncie, IN, USA \\
		\begin{cvlist}
			\item[] M.S. Physics,  July 2009
			\begin{cvlistb}
				\item Thesis Title:  ``In Search of Red Dwarf Stars: Application of Three-Color Photometric Techniques'' 
				\item Advisor:  Dr. Thomas Robertson, Department Chairperson
			\end{cvlistb}
			\item[] B.S. Applied Physics, July 2007
		\end{cvlist}
		
		\section*{\sc Research Interests}
		Condensed matter astrophysics, structure in the white dwarf luminosity function, neutron star crusts, optical observations, variations in the M dwarf luminosity function for varying galactic latitude, structure of Milky Way Galaxy spiral arms, data mining
		
		\section*{\sc Research Experience}
		{\bf Indiana University:} (Supervisor - Charles Horowitz) \\
		Investigated the quantum effects in dense plasmas in compact stellar systems such as white dwarfs and neutron stars. Determined ionic diffusion coefficients and melting temperatures through the use of molecular dynamics simulations.

		{\bf Ball State University:} (Supervisor - Thomas Robertson) \\
		Helped in the development of a photometric system in which to distinguish M dwarf stars from M giants. Attempted to find local variations in the M dwarf luminosity function as a function of galactic latitude.

		{\bf Ball State University:} (Supervisor - Thomas Robertson) \\
		Performed multiple online catalog searches as part of a proper motion survey in the search for M dwarf candidates. 

		\section*{\sc Data Analysis \& Observing Experience}
		{\bf Scientific Computing:} Several years experience with compiled (C++) programming. Performed Molecular Dynamics simulations of up to $\sim$16,000 ions to simulate the internal conditions of white dwarf stars and the crusts of neutron stars.

		{\bf Optical Photometry:} Prepared and executed imaging programs for calibrated photometry field stars. Performed reduction and analysis of imaging data for point sources.

		{\bf Programming:} Several years experience with Image Reduction and Analysis Facility (IRAF) and ds9.

		\section*{\sc Honors and Awards} 
		\begin{cvlistb}
			\item  Indiana Space Grant Consortium Graduate Fellowship, 2008
			\item  Indiana Space Grant Consortium Scholarship, 2006
			\item  Recipient of the Keys-Litten-Smith and Sigma Xi Outstanding Graduate Poster Award, 2009
			\item  Inducted into Sigma Pi Sigma, 2006
			\item  Inducted into National Scholars Honor Society, 2006
		\end{cvlistb}
			
		\section*{\sc Conferences \& Workshops}
		\begin{cvlistb}
			\item  WIYN telescope ODI/PPA workshop, June 2011
			\item  41st Annual HASTI Conference, February 2011
			\item  212th meeting of the American Astronomical Society, June 2008
			\item  211th meeting of the American Astronomical Society, January 2008 
		\end{cvlistb}
		
		\section*{\sc Teaching Experience}
		{\bf Indiana University} \\
		{\em Instructor -- Astro 105 } \hfill {\bf May - June 2011}\\
		Designed and taught my own introductory course for non-majors focusing on stars, galaxies, and cosmology. \\
		\vspace{3mm}
		{\em Instructor -- Astro 100 } \hfill {\bf May - June 2010}\\
		Designed and taught my own introductory course for non-majors focusing on the solar system.	\\
		\vspace{3mm}
		{\em Associate Instructor -- Astro 100 } \hfill {\bf January - May 2011}\\
		{\em Associate Instructor -- Astro 100 } \hfill {\bf January - May 2010}\\
		Introductory course on the Solar System for non-majors and general astronomy for majors.  Assisted with lectures and demos, taught when professor was absent, held office hours, and graded homework.
		
		{\em Associate Instructor -- Astro 105 } \hfill {\bf August - December 2010}\\
		Introductory course on stars and galaxies for non-majors and general astronomy for majors.  Assisted with lectures and demos, taught when professor was absent, held office hours, and graded homework.

		{\em Associate Instructor -- Astro 305 } \hfill {\bf September - December 2009}\\
		Observational techniques class for majors.  Assisted with training students on using university-owned telescopes, helped with night-time observing sessions, and held office hours.\\
		\vspace{5mm}
		{\bf Ivy Tech Community College}\\
				{\em Adjunct Faculty -- Astronomy 101} \hfill {\bf June - August 2011}\\
		Taught four distance learning sections of an online introductory astronomy course about the solar system. Included an online laboratory component.\\
		\vspace{3mm}
{\em Adjunct Faculty -- Physics 101} \hfill {\bf January - May 2011}\\
		Designed and implemented an introductory algebra based physics course that fulfilled state standards. Lead laboratory experiments correlating to course material.\\
		
		\vspace{5mm}
		{\bf Ball State University}\\
		{\em Teaching Assistant -- Physics 101 \& 110} \hfill {\bf August 2007 - July 2008}\\
		Physics courses for elementary education majors and for general university requirements respectively. Graded homework and exams. Maintained office hours for students to receive tutoring as needed.

		{\em Lab Assistant -- Physics 101 \& 110} \hfill {\bf January 2006 - June 2007}\\
		Physics courses for elementary education majors and for general university requirements respectively. Set up and guide laboratory experiments based on given curriculum. Assist students with lab-related work outside of class hours.

		\section*{\sc Public Outreach Activities}
		Participated in educational activities at both Ball State University and Indiana University.  Such activities include tours of the campus observatories, the yearly IU Physics \& Astronomy Open House, demonstrations for the World Year of Physics, and running planetarium events which were typically for elementary through high school students. Judge at the 2009 East Central Indiana Region Science Fair and twice judged at the Intel ISEF science fair.
		
		\section*{\sc Service}
		\begin{cvlistb}
		\item Member of the Swain West Green Team as an effort to make Indiana University more energy efficient and self sustainable\hfill {\bf March 2010 - August 2011}
		\item Planetarium assistant, Ball State University \hfill {\bf August 2008 - July 2009}
		\item Representative of the Physics and Astronomy Department on the Dean Advisory Committee at Ball State University\hfill {\bf January 2007 - May 2007}
		\end{cvlistb}

		\section*{\sc Conference Presentations}
		\begin{cvbib}
			\item[] ``In Search of Red Dwarf Stars: Application of Three Color Photometric Techniques".  J. Mason \& T. H. Robertson, Poster AAS 212.11.03, 2008.
			\item[] ``Luminosity Classification of Potential M Dwarf Stars Selected Using 2MASS and Tycho2 Data". J. Mason, N. Humphrey, A. Briggs, A. Parrell, \& T. H. Robertson, Poster AA2 2011.163.03, 2008.
			\item[] ``Assessment Three Color Photometric Techniques".  J. Mason \& T. H. Robertson, Butler University's 20th Annual Undergraduate Research Conference, 2007.
			
		\end{cvbib}
	\end{vitae}

\end{document}